\documentclass[usenatbib]{mn2e}
\usepackage{graphicx}
\usepackage{deluxetable}

\def\gax{\mathrel{\raise.3ex\hbox{$>$}\mkern-14mu\lower0.6ex\hbox{$\sim$}}}
\def\lax{\mathrel{\raise.3ex\hbox{$<$}\mkern-14mu\lower0.6ex\hbox{$\sim$}}}
\def\gtorder{\mathrel{\raise.3ex\hbox{$>$}\mkern-14mu
             \lower0.6ex\hbox{$\sim$}}}
\def\ltorder{\mathrel{\raise.3ex\hbox{$<$}\mkern-14mu
             \lower0.6ex\hbox{$\sim$}}}

\begin{document}

\title[Dust Formation in the Presence of Photons I]
   {Dust Formation in the Presence of Photons I: Evaporation Rates for Small Dust Grains}

\author[C.~S. Kochanek]{ C.~S. Kochanek$^{1,2}$  \\
  $^{1}$ Department of Astronomy, The Ohio State University, 140 West 18th Avenue, Columbus OH 43210 \\
  $^{2}$ Center for Cosmology and AstroParticle Physics, The Ohio State University,
    191 W. Woodruff Avenue, Columbus OH 43210
   }

\maketitle

\begin{abstract}
The temperature of newly forming dust is controlled by the radiation
field. As dust forms around stars, stellar transients, quasars or supernovae,
the grains must grow through a regime where they are stochastically
heated by individual photons.  Since evaporation rates increase
exponentially with temperature while cooling times decrease only
as a power law, the evaporation rates for these small grains are
dominated by the temperature spikes.  We calculate effective 
evaporation temperatures for a broad range of input spectra
that are encapsulated in a series of simple interpolation 
formulae for both graphitic and silicate grains.  These can 
be easily used to first determine if dust formation is possible
and then to estimate the radius or time at which it commences
for a broad range of radiation environments. With these
additional physical effects, very small grains may form earlier 
than in standard models of AGB winds.  Even for very high mass
loss rates, the hottest stars that can form dust are G and F
stars particularly in the case of silicate dusts.  For hotter
stars, the higher fluxes of ultraviolet photons prevent dust 
formation. Thus,  
episodic dust formation by OH/IR stars and LBVs is primarily
driven by fluctuations in their apparent temperatures rather
than changes in luminosity or mass loss rates.    
\end{abstract}


\section{Introduction}
\label{sec:introduction}

Dust formation and destruction is crucial to interpreting a broad range of astrophysical
phenomena including stellar winds, stellar transients, supernovae and quasars.  In the 
laboratory, particle formation and growth is simply a collisional process because both 
the temperature and the growth rates are determined by collisions.  Radiation is
unimportant because the densities are high ($n \sim 10^{19}$~cm$^{-3}$) and the 
(radiation) temperatures are low ($T \ltorder 2000$~K).  In most astronomical contexts, however, 
the reverse is true -- the densities are low ($n \ltorder 10^{10}$~cm$^{-3}$)
and the radiation temperatures are high ($T \gtorder 2000$~K). In these
circumstances, particles grow by collisions but the grain
temperatures, and hence the evaporation rates, are controlled by the radiation field.  

Dust formation can be viewed as having two phases, nucleation and 
growth (see, e.g., the reviews by \citealt{Salpeter1977}, \citealt{Gail2013}).  The distinction is 
that there is generally a ``critical cluster'' 
size or number ($a_{min}$, $N_{min}$), below which a grain will be
unstable to returning to its simpler starting components (``monomers'') 
and above which it becomes
a stable grain that will continue to grow as it collides with more 
``monomers'' or other grains.   Nucleation theory predicts the
rate of formation of critical clusters based on the thermodynamics of the
gas, but it does so poorly, and nucleation rates at fixed temperature are
generally exponentially uncertain (e.g., \citealt{Wolk2002}).  This is exacerbated in astrophysical
flows by the very long collision time scales (e.g., \citealt{Donn1985}).
In expanding flows, however, the temperature steadily drops
and the supersaturation of the vapor phase steadily rises, making a
failure to nucleate exponentially unlikely even if the rate at any
given temperature is very uncertain.  Once grains nucleate and start to grow, 
their final sizes are determined by kinetics.  

As grains try to grow, they must pass through a phase where the number
of constituent particles is small, $10 \ltorder N \ltorder 100$, and
the specific heat of the grain is low.  Such small grains can be 
stochastically heated to very high temperatures by individual soft
ultraviolet (UV) photons, as was originally discussed in the context of
the stochastic heating of grains in the interstellar
medium (e.g., \citealt{Sellgren1984},
\citealt{Draine1985}, \citealt{Desert1986}, 
\citealt{Guhathakurta1989}), the destruction of small
grains and polycyclic aromatic hydrocarbons (PAH) near active 
galactic nuclei (AGN, e.g., \citealt{Voit1991}, \citealt{Voit1992}),
and dust formation by novae (\citealt{Johnson1993}).
Both the shorter wavelength fluxes and the evaporation rates of small
grains are dominated by the short duration temperature spikes
because the emission or evaporation rises exponentially with
grain temperature while the time spent at the highest temperatures
drops only as a power-law.

Stochastic heating is rarely mentioned in the context of dust
formation.  For example, many studies of
dust formation by supernovae ignore radiative heating entirely
and set the grain temperature to that of the gas
(e.g., \citealt{Kozasa1991}, \citealt{Todini2001}, \citealt{Nozawa2003}, 
\citealt{Bianchi2007}).  Others, such as \cite{Dwek1988} for SN,
\cite{Clayton1976} for novae and \cite{Lefevre1979} for stars and 
novae set the grain temperature to be in equilibrium
with the absorbed flux.  A few studies such as \cite{Guhathakurta1989}
for dust in the ISM, \cite{Johnson1993} 
for dust formation in novae, \cite{Cherchneff2009} for the effects 
of hard photons on the pre-cursor chemistry to dust formation in SN, and 
\cite{Keith2011} for the growth of small carbonaceous grains have
considered the role of stochastic heating.   

In most astrophysical contexts, gas collisions are unimportant
to the thermodynamic state of a grain.\footnote{The critical density
for collisions to dominate over radiation given grain and gas temperatures
of $T_{g3}$ and $T_{gas3}$ in units of $1000$~K is
\begin{equation}
    n \gtorder 10^{15} { Q(T_g) T_{g3}^4 \over |T_{gas3}-T_{g3}| T_{gas3}^{1/2}}~\hbox{cm}^{-3}. 
\end{equation}
}
The grain heats and cools
by absorbing and emitting photons.  Small grains are also thermally
isolated in the sense that the grain experiences no collisions with
gas particles over the time scale needed for the grain to cool after
absorbing a photon (\citealt{Guhathakurta1989}, \citealt{Johnson1993}).  This makes dust formation extremely sensitive
to the hardness of the radiation field, and provides a natural explanation
of why dust fails to form around hot stars even for very high mass
loss rates (\citealt{Kochanek2011}).  Around the very cool stars that
are the focus of most studies of dust formation (e.g., \citealt{Rowan1983},
\citealt{Gauger1990}, or more recently, \citealt{Nanni2013} and
\citealt{Ventura2014}), the 
effect becomes relatively unimportant because the stellar
temperatures are too low to drive strong stochastic heating.
However, for dust formation around hotter stars, stellar transients, 
novae, supernovae and AGN, these radiative 
effects cannot be neglected.

Without stochastic heating, dust formation in outflows seems
almost inevitable provided there is a mechanism to initiate
the outflow (see, e.g., the review by \citealt{Lafon1991}).  
For simplicity, consider an $N$ particle 
grain near a source of luminosity $L_*$ emitting
photons of energy $E_\gamma= k T_\gamma$ and ignoring Planck 
factors.
In equilibrium, the grain has a black body temperature of 
$T_{bb}=(L/16\pi\sigma r^2)^{1/4}$ when it is a distance $r$
from the source.  Evaporation rates depend exponentially on
temperature, $\propto \exp(-T_c/T_{bb})$, where $k T_c$ is the 
binding energy of particles to the grain (e.g., \citealt{Lefevre1979},
\citealt{Guhathakurta1989}), so evaporation rates
drop exponentially with distance since $T_c/T_{bb}\propto r^{1/2}L^{-1/4}$.  
Collisional growth rates are proportional to density, so in
an expanding (wind) flow the growth rate also declines with distance
but only as a power law, $\propto r^{-2}$.  At some radius
that is only logarithmically dependent on the spectrum or the 
density, the collisional growth rate exceeds the evaporation
rate and the grain will begin to grow unless the densities are
so low that there is no probability of experiencing any 
collisions.  All sufficiently dense flows should form dust.  

Stochastic heating fundamentally changes the scaling of the
evaporation rate with distance.  Suppose that a grain containing
$N$ particles has 
time to completely cool between photon absorptions.  When the 
grain absorbs a photon, it is heated by temperature
$\Delta T = T_\gamma/c_V N$ where $c_V \sim 2$-$3$ is the
specific heat of the grain.  If $\Delta T \gtorder T_{bb}$, all evaporation
occurs at the temperature $\Delta T$ set by the photon energy
rather than the black body temperature $T_{bb}$ set by the
photon flux.  However, the radiative cooling time scale,    
$t_{cool}= c_V N k \Delta T/ 4 \pi a^2 \sigma \Delta T^4$,
for a grain of radius $a$ is short compared to the
rate of photon absorptions, $R = L\pi a^2 /4 \pi r^2 E_\gamma$, 
leading to an average evaporation rate of 
$\propto \exp(-T_c/\Delta T) t_{cool} R = \exp(-T_c/\Delta T) L/ r^2 \Delta T^4$. 
The term in the exponential is constant,
so the evaporation rate now drops with the same $r^{-2}$ radial power law as
the collisional growth rates and distance does not affect the balance
between evaporation and growth.  
Only modest photon energies are needed to drive this process
for small grains, since photons of energy $E_\gamma \simeq 0.2 c_V N$~eV
will heat a grain by $2000$~K and all but the coldest stars and
transients produce large numbers of such photons.

Our original interest in this question comes from the properties
of ejecta around Luminous Blue Variables (LBVs), massive hot stars
with dense winds ($\dot{M}\sim 10^{-5}$ to $10^{-4}M_\odot$/year) 
that are not presently forming dust yet are frequently surrounded
by massive dusty shells of material (see the reviews by 
\citealt{Humphreys1994}, \citealt{Vink2012}).  The most recent
example is $\eta$~Car, but there are some 10(s) of examples in
the Galaxy.   In \cite{Kochanek2011} we argued that this could
be explained by stochastic heating of small grains.  Despite
the density of their present day winds, the abundance of soft
ultraviolet photons prevents dust formation.  In eruption, the
wind becomes optically thick, leading to a pseudo-photosphere 
that is much cooler (\citealt{Davidson1987}), and this allows
dust to form.  For dust formation, this change in radiation 
temperature is more important than the increase in wind density,
although the higher wind density ($\gtorder 10^{-3}M_\odot$/year)
required to produce the pseudo-photosphere certainly accelerates
the growth of the grains.  Thus, the effects of stochastic
heating provide a natural explanation for hot stars surrounded
by young ($<10^4$~year), dusty shells.
 
More generally, the nature of the radiation field is likely 
important for dust formation in all astrophysical scenarios
where the radiation temperature is high compared to the dust
evaporation temperature.  This includes LBVs, OH/IR stars, 
novae, supernovae, other stellar transients (e.g., supernova ``impostors'', 
and pre-super mass loss events) and quasars.  
In \cite{Kochanek2011} we simply compared the collision rates
to the soft UV photon absorption rate.  Here, in \S\ref{sec:form}
we explicitly calculate the stochastic heating of small grains to
obtain interpolation formulae that can be used in any context.
We discuss the differences between various treatments of the
radiative heating of small grains in \S\ref{sec:results}.
In \S\ref{sec:winds} we show how these results apply to 
steady winds as a function of stellar temperature and mass
loss rates.  In \S\ref{sec:discussion} we discuss a few of
the implications for dust formation in near AGB stars,
OH/IR stars and long-lived LBV eruptions such as the Great
Eruption of $\eta$ Carinae.  Detailed discussions of dust formation 
in shorter transients, supernovae and quasars will follow
separately. 

\section{Dust Formation }
\label{sec:form}

We are generally considering dust formation near a star of mass $M_*=10M_{*10}M_\odot$,
luminosity $L_* = 10^4 L_{*4} L_\odot$, temperature $T_*=10^4 T_{*4}$~K and
radius $R_*=L_*/4\pi \sigma T_*^4$, forming dust in a wind with
mass loss rate $\dot{M} = 10^{-4} \dot{M}_4 M_\odot$/year and velocity
$v_w = 10 v_{w10}$~km/s or scaled by the stellar escape speed
\begin{equation}
     v_w = \beta \sqrt{ 2 G M_* \over R_*} 
        \simeq  338\beta M_{*10}^{1/2} T_{*4} L_{*4}^{-1/4}~\hbox{km/s}.
       \label{eqn:vesc}
\end{equation}
For a wind we use the simple constant velocity density profile
\begin{equation}
     \rho = { \dot{M} \over 4 \pi v_w r^2 }
\end{equation}
of which mass fraction $X = 0.01 X_2$ is comprised of the condensible
species ($X$ can be much higher for (super)nova ejecta).    
Dust forms in grains of radius $a$ containing $N=4\pi a^3 \rho_b/3 m_0$
particles where $\rho_b$ is the bulk density and $m_0$ is the (average) mass
of a monomer.  Following \cite{Guhathakurta1989}, we use $m_0=12 m_p$ ($20m_p$) 
and a bulk density of $\rho_{bulk}=2.2$ ($3.5$)~g~cm$^{-3}$ for graphitic 
(silicate grains).  Note that the ratio $\rho_{bulk}/m_0$ is essentially
identical for the two compositions, so $ a = 0.00013 N^{1/3}\mu$m in
both cases.  We will be discussing very small ($10 \leq N \leq 100$) ``grains'',
and we will discuss the applicability of using bulk properties as we proceed.
We use the dust cross sections and Planck averaged cross sections from 
\cite{Draine1984} and \cite{Laor1993} extending them below $0.001\mu$m
with $Q\propto a$ and above $10\mu$m with $Q$ constant.

Our objective is to determine the conditions under which a small grain
containing $N$ monomers can grow given the density and radiation environment.
Growth is set by the balance between collisional gains and evaporative losses.
In these early phases, we can focus on growth by collision with monomers, so 
the growth rate is
\begin{equation}
     \left( { dN \over d t }\right)_c = R_c = { \alpha \pi a^2 X \rho v_c \over m_0}
      \label{eqn:coll}
\end{equation}
where $v_c$ is the collision velocity and $\alpha \leq 1$ is the ``sticking probability''
(e.g., \citealt{Kwok1975}, \citealt{Deguchi1980}, \citealt{Gail1984}, \citealt{Gail1988}).  
When a monomer condenses onto a grain, the grain is heated by the energy 
of condensation, increasing the probability of then evaporating a monomer.
The sticking probability can be used to model the resulting, net condensation rate and
it becomes smaller for smaller grains (\cite{Johnson1993} estimate that 
$\alpha \simeq 0.6$ for $N=10$ but only $10^{-4}$ for $N=3$) because
small grains have fewer vibrational degrees of freedom.  We will only
consider larger ($N>10$) grains and we will consider a model in which
the effects of condensation and evaporation are included when evolving
the grain enthalpy.  Thus,  we can generally set $\alpha=1$ since small 
changes in $\alpha$ are also unimportant compared to the potential range of 
wind densities.  
Associated with any flow at radius $R$, there is also an expansion time, 
$t_e=R/v_w$.  Growth is kinetically limited, in the sense that there are no collisions, 
if the collision rate is less than the expansion rate, $(dN/dt)_c < t_e^{-1}$.

The collision velocity $v_c$ is a combination of gas temperature, drift velocities 
and any turbulent
velocities.  Since the gas is cold, $T_{gas}=1000T_{gas3}$~K, and the monomers are 
relatively heavy, the thermal velocities of the monomers are low,
$v_{th} \simeq 0.8 T_{gas3}^{1/2}$~km/s (one velocity component,
carbon).  The thermal velocities of the grains are then still lower by $N^{-1/2}$.
The grains also drift relative to the gas at velocity  
\begin{equation}
    v_d = \left( { Q_{rp} (T_*) L_* \over 4 \pi r^2 \rho c } \right)^{1/2},
\end{equation}
which is the balance between radiation pressure accelerating the grains and 
collisional drag from the gas (e.g., \citealt{Netzer1993}).  The radiation pressure Planck factor, $Q_{rp}$, is 
roughly equal to the absorption Planck factor $\langle Q_{abs} \rangle$ 
we use below.  If we substitute the density of a wind, then
\begin{equation}
      v_d \simeq 0.3 \beta^{1/2} L_{*4}^{3/8} M_{*10}^{1/4} \hat{Q}_{rp}^{1/2} 
          T_{*4}^{1/2} \dot{M}_4^{-1/2} N^{1/6}~\hbox{km/s}
\end{equation}
where $Q = \hat{Q}(a/\mu{m})$ and $\hat{Q}_{rp}(T_*)$ is small
for cool stars ($0.1$-$1$) and larger for hot stars ($10^2$).\footnote{For $2600 < T_* < 50000$~K 
and small grains, they can
be approximated as $\ln \hat{Q} \simeq 2.969+1.578 x -0.176 x^2$
for graphite and $0.600+2.925 x+1.141 x^2$ ($T_*<10^4$~K) and 
$0.600+4.378 x-1.118 x^2$ ($T_*>10^4$~K) for silicate where $x=\ln(T_*/10^4)$.}
If $v_d \gtorder v_{crit} \simeq 50$~km/s, sputtering
by collisions with the carrier gas will destroy the grain (e.g., \citealt{Draine1995}).  In
theory this is possible for very low $\dot{M}$, but at such low wind
densities dust growth is already impossible due to kinetics.
Like the sticking probability, the dynamic range of $v_c$ is small 
compared to that of the density, so we will
simply scale the collision velocity by $v_c = 1 v_{c1}$~km/s.  
  

The formation radius $R_f$ represents the point at which the grain can begin to grow
faster than it evaporates, where the grain temperature $T_g$ determines the 
evaporation rate, 
\begin{equation}
     \left( { dN \over dt }\right)_e = R_e = - 4\pi a^2 R(N,T_g) S(N,T_g). 
        \label{eqn:evap}
\end{equation}
Here $R(N,T_g)$ is the evaporation rate per unit area in thermal equilibrium
and $S(N,T_g)$ accounts for the suppression of thermal fluctuations if the 
grain is thermally isolated from the surrounding gas 
in the sense that radiative time scales are shorter than collisional time scales
(\citealt{Guhathakurta1989}).
The evaporation rates are modeled as 
\begin{equation}
    R(N+1,T) = A \exp\left[-B/T \right] 
   \label{eqn:evap0}
\end{equation} 
with
\begin{eqnarray}
     A &= &4.6 \times 10^{30} \alpha~\hbox{cm}^{-2}~\hbox{s}^{-1}
    \quad\hbox{and} \nonumber \\
     B &= &81200-20000(N-1)^{-1/3}~\hbox{K},
\end{eqnarray} 
for graphite and 
\begin{eqnarray}
     A &= &4.9 \times 10^{31} \alpha~\hbox{cm}^{-2}~\hbox{s}^{-1}
    \quad\hbox{and}\nonumber \\
     B &= &68100-20000(N-1)^{-1/3}~\hbox{K},
\end{eqnarray} 
for silicates where $\alpha$ is the ``sticking probability'' 
(\citealt{Guhathakurta1989}).  Since $\alpha$ appears
in both the collision and the evaporation rates, we simply set
$\alpha =1 $.  The suppression factor
\begin{equation}
   S(N,T) = \left( { 1 + \gamma \over \gamma }\right)^b 
                { \Gamma[\gamma f +1 ]\Gamma[\gamma f + f - b]\over
                         \Gamma[ \gamma f - b +1] \Gamma[\gamma f + f ] }
\end{equation}
reduces the rate below that for thermal equilibrium ($R(N,T)$)
by the probability of having enough vibrational quanta
in the bond holding a surface atom onto the grain given that the number of vibrational quanta
is fixed if the grain is in isolation.  The grain has $f = 3 N -6$ vibrational degrees of freedom 
with an (assumed) common vibrational energy $\hbar \omega_0 = 0.75 k \Theta$ of the Debye
temperature ($\Theta=420$~K for graphite, $470$~K for silicates).  The quantity $\gamma = U/f \hbar \omega_0$
is the mean number of vibrational quanta per degree of freedom given the internal energy 
of the grain and $b = k B/\hbar \omega_0$.  If $\gamma f - b + 1 \leq 0$, then the evaporation
is completely suppressed, and this effectively corresponds to the requirement that
$U > k B$ before any evaporation is possible.  As discussed in \S1, the grains are
essentially always thermally isolated from the carrier gas on the time scale for 
cooling after absorbing a photon. 
Again following \cite{Guhathakurta1989}, the enthalpy of an $N$ atom graphitic grain can be 
approximated as
\begin{equation}
     U_N = { 4.15 \times 10^{-22} (N-2) T^{3.3}~\hbox{ergs/atom} \over
    1 + 6.51 \times 10^{-3} T + 1.5 \times 10^{-6} T^2 + 8.3 \times 10^{-7} T^{2.3} },
\end{equation}
while for silicates 
\begin{eqnarray}
   &U_N  &= (N-2) (10^{-21}~\hbox{ergs/atom})\times \\ 
        &&\left\lbrace 
         \begin{array}{ll}
              4.43 T^3                                                       &T<50 \\
              7.33 \times 10^5 (T/50)^{2.3} - 1.80\times 10^5           &50 < T < 150 \\
              1.23 \times 10^7 (T/150)^{1.68} - 3.27 \times 10^6        &150 < T < 500 \\
              1.62 \times 10^8 (T/500) - 7.23 \times 10^7               &500 < T.
         \end{array}
         \right. \nonumber
\end{eqnarray}
The energy required to heat a grain to $1500$~K is roughly $0.25 (N-2)$~eV in both cases,
so the flux of $1$-$10$~eV photons controls the importance of stochastic heating, particularly
since hydrogen will generally absorb $>13.6$~eV photons.

The simplest way to consider particle formation is to view it as completely suppressed
as long as the evaporation rate (Equation~\ref{eqn:evap}) exceeds the collisional
growth rate (Equation~\ref{eqn:coll}).  Without stochastic heating, the evaporation
rate depends on the nature of the incident spectrum because the black body equilibrium 
temperature  
\begin{equation}
             { L \over 4 \pi r^2 } = 4 \sigma T_{bb}^4 
     \label{eqn:bbtemp}
\end{equation}
differs from the Planck equilibrium temperature
\begin{equation}
             { L \langle Q_{abs} \rangle \over 4 \pi r^2 } = 4 \sigma T_g^4 Q_{em}(T_g)
     \label{eqn:equil}
\end{equation}
where
$\langle Q_{abs} \rangle = \int F_\nu Q_{abs}(\nu) d\nu/\int F_\nu d\nu$ depends 
on the incident radiation spectrum and $Q_{em}(T_g)$ on the grain temperature.
Generally $\langle Q_{abs} \rangle$ increases with
the radiation temperature and is larger than $Q_{em}(T_g)$, so 
the grain is hotter than predicted from the black body temperature.  
Equating the collisional growth (Equation~\ref{eqn:coll}) to the evaporation rate (Equation~\ref{eqn:evap}),
a grain can grow if  
\begin{eqnarray}
       T_g &<  &B \left[ \ln \left( { 4 A m_0 S \over X \rho v_c } \right) \right]^{-1} \\
           &\simeq  &B 
     \left[ 50 + \ln\left( { S A \over 10^{31}} { 10^5 \over n_{cond} } { \hbox{km/s}\over v_c} \right) \right]^{-1} 
         \nonumber
        \label{eqn:tcrit0}
\end{eqnarray}  
where the number density of the condensible species is $n_{cond}=X\rho/m_0$. This
basically leads to a characteristic temperature of order $B/50\sim 1500$~K because reasonable
changes in the variables entering the logarithm do not produce changes that are 
significant fractions of the leading constant.  For example raising the collisional
growth rate by $10^5$ (e.g., raising $n_{cond}$ from $10^5$ to $10^{10}$~cm$^{-3}$)
only changes the limit from $T_g<B/50$ to $B/38$.   

In equilibrium, the thermal evolution is controlled by two effects.  The
first, familiar effect is radiation balance
\begin{equation}
          { dU \over dt } = - 4 \pi a^2 \sigma T_g^4 Q_{em}(T_g) + { L \langle Q_{abs} \rangle \pi a^2 \over 4 \pi r^2}.
  \label{eqn:evolve}
\end{equation}
Here the second term is the mean radiative heating, but we can also treat it 
stochastically in terms of individual photons.  We must also,
however, account for the energy associated with adding or evaporating a 
particle from the grain (e.g., \citealt{Waxman2000}),
\begin{equation}
  { dU \over dt } = (R_e+R_c)k B 
  \label{eqn:evolve2}
\end{equation}
where $R_c=(dN/dt)_c >0$ and $R_e=(dN/dt)_e<0$ are the 
condensation (Equation~\ref{eqn:coll}) and 
evaporation (Equation~\ref{eqn:evap}) rates, respectively.
Since we are searching for the point where evaporation and growth
are in balance, the average of this term is zero.  However,
the heating from particle addition is uncorrelated with grain
temperature, while evaporation will preferentially occur when
the temperature is high.  Thus, high temperature peaks cool
faster when we include these terms and, as you add more heat,
the evaporation rate increases rather than the mean temperature.
These terms will be treated stochastically.  As discussed above,
it is easily shown that collisions with the carrier gas are 
generally unimportant to the thermal balance.  

Operationally, we carried out the calculations as follows.  We treated
photons with energy $<0.1$~eV as contributing a steady heating term
in Equation~\ref{eqn:evolve} and computed the cooling curve of a 
grain $U(t)$ and temperature $T(t)$ in the absence of any stochastic event.  
We also computed the total number of particles $\Delta N(t)$ 
evaporated.  Given these curves, the remainder of the calculation is simply
an iterative procedure with no need for further numerical integration.  From the
photon spectrum, we can determine the rate of discrete photon absorptions.
We work with normalized spectra, $b_\nu = B_\nu/\int B_\nu d\nu$, so 
the flux of photons of energy $h\nu$ at any radius is $F_\nu = L b_\nu/4 \pi r^2$.  
Given this photon flux, the rate at which photons are absorbed by the grain
is $R_\gamma = \pi a^2 \int F_\nu Q_{abs}(\nu) (h\nu)^{-1} d\nu$.
The photon energy $E_\gamma$ is found by randomly sampling the
distribution $\nu^{-1} b_\nu Q(\nu)$.  We also have the current evaporation rate
$R_e$ and the average capture rate $R_c = \langle R_e \rangle$ which
are set to be equal because we are solving for the conditions where they
just balance.  

The rate of events is then $R=R_\gamma + R_e + R_c$, so the time until
the next event is $\Delta t = -R^{-1}\log(1-P)$ where $P$ is a uniform 
random deviate.  Given the current grain enthalpy, $U(t)$, we can
follow the tabulated cooling curve to $U(t+\Delta t)$ and the
contribution of this time interval to the overall amount of
evaporation.  Based on the relative
rates we can randomly select the next event, and update the enthalpy
by adding the energy of the absorbed photon, $U(t+\Delta t) + E_\gamma$,
cool by evaporating a monomer, $U(t+\Delta t) - k B$, or
heat by capturing a monomer, $U(t+\Delta t) + k B$.  We start
the grain at its Planck equilibrium temperature and use a ``break in''
period to forget this initial condition and bring the capture
rate to equal the mean evaporation rate.  From the end of the
break in period to the end of the calculation we total the 
evaporations and then divide by the elapsed time to obtain
the mean evaporation rate $\langle dN/dt\rangle_e$.  Typically
we followed $10^7$ events.  

We tracked five different cases for comparisons:
\begin{itemize}
\item  Case 1 (Planck equilibrium): thermally coupled grains at the equilibrium temperature and only photons 
      (Equation~\ref{eqn:evap} with $S\equiv 1$ plus Equation~\ref{eqn:equil} and no Equation~\ref{eqn:evolve2});
\item  Case 2: thermally isolated grains at the equilibrium temperature and only photons 
      (Equation~\ref{eqn:evap} plus Equation~\ref{eqn:equil} and no Equation~\ref{eqn:evolve2});
\item  Case 3: stochastically heated grains that are thermally coupled and only photons 
      (Equation~\ref{eqn:evap} with $S\equiv1$ plus Equation~\ref{eqn:evolve} and no Equation~\ref{eqn:evolve2}); 
\item  Case 4: stochastically heated grains that are thermally isolated and only photons 
      (Equation~\ref{eqn:evap} plus Equation~\ref{eqn:evolve}, and no Equation~\ref{eqn:evolve2}); and
\item  Case 5: stochastically heated grains that are thermally isolated with evaporation and capture 
       (Equation~\ref{eqn:evap} plus Equation~\ref{eqn:evolve} and Equation~\ref{eqn:evolve2}).
\end{itemize}
We can only include the heating and cooling effects of evaporation and capture (Equation~\ref{eqn:evolve2})
as a modification
to the models including the suppression factor $S$, as this is the term the prevents evaporation
when the grain temperature is too low and guarantees that $U(t+\Delta t) - k B > 0$.
In the next section we illustrate the effects of these various cases, and then in the later sections 
we will narrow our discussion to comparisons of Case 1 (Planck equilibrium) and Case 4 (stochastically
heated, thermally isolated grains).
We treat Case 1 as the standard for comparison because almost all studies that
set grain temperatures based on the radiation field assume Planck equilibrium temperatures.

The balance between growth and evaporation can be described by two
characteristic temperatures.  The radiation is characterized by
the grain black body temperature, $T_{bb} = (L/16 \pi \sigma r^2)^{1/4}$ 
(Equation~\ref{eqn:bbtemp}), which is independent of any dust 
properties and simply represents the photon energy flux.\footnote{This could also include the geometrical dilution
factor, but we do not include it here for simplicity.}   
Almost all the properties of the dust and the
spectrum can be encapsulated using an effective evaporation temperature of 
\begin{equation}
    T_e = - B^{-1} \ln \left( {-\langle dN/dt \rangle_e \over  4\pi a^2 A } \right),
    \label{eqn:tevap}
\end{equation}
found by inverting Equation~\ref{eqn:evap} with $S\equiv 1$. 
For any given spectrum and model for the physics describing the heating
of the grains, we can calculate the mapping between the radiative flux
characterized by $T_{bb}$
and the evaporation rate, characterized by the effective
evaporation temperature, $T_e(T_{bb})$.  The gas density implies a
critical effective evaporation temperature for growth
which is just Equation~\ref{eqn:tcrit0} with $S \equiv 1$.  For
graphite this becomes
\begin{eqnarray}
       T_{crit} &= &\left[1460 - 360 (N-1)^{-1/3}\right] \times \\
         & &\left[1 - \left(\ln n_5 v_{c1} X_2\right)/55.8)\right]^{-1}~\hbox{K} \nonumber
      \label{eqn:tcrit1}
\end{eqnarray}
and for silicates it becomes
\begin{eqnarray}
       T_{crit} &= &\left[1120 - 330 (N-1)^{-1/3}\right]  \times \\
       & &\left[1 - \left(\ln n_5 v_{c1} X_2\right)/60.9)\right]^{-1}~\hbox{K} \nonumber
      \label{eqn:tcrit2}
\end{eqnarray}
where the density is related to the hydrogen number density $n$ by 
$\rho= 4 n m_p/3$ for roughly Solar abundances.  The condition
for growth is then simply that $T_{crit} > T_e (T_{bb})$.  

If we carry out our simple calculation from \S1 for heating grains with mono-energetic
photons more carefully, assuming that the grains have a mean temperature $\langle T \rangle$
and are heated to a peak temperature of $T_{peak} = \langle T \rangle + \Delta T$,
we find that the effective evaporation temperature is
\begin{equation}
        T_e = T_{peak} 
    \left[ 1 + { T_{peak} \over B} 
    \ln \left( { T_{peak}^3 \Delta T \over S(T_{peak}) \langle T \rangle^4 } \right)\right]^{-1}.
      \label{eqn:model}
\end{equation}
Because $T_{peak}/B \sim 10^{-2}$ multiplies a logarithm, the effective
evaporation temperature is essentially $T_e \simeq T_{peak}$.  At
large distances $T_{peak} \rightarrow \Delta T$ and for high energy
photons ($E_\gamma \gtorder kB$) we recover the argument in \S1 that the evaporation rate 
does not cut off exponentially at large distances from the source.
For low energy photons, the suppression factor becomes exponentially
small, the evaporation rate is cut off exponentially at large
distances, and we return to the case where stochastic heating is
unimportant.  In the limit that the grain cools completely between photons, we can
calculate the effective evaporation temperature to be
\begin{equation}
      T_e =  B \left[ 4 \ln (T_0/T_{bb}) \right]^{-1}
       \label{eqn:limit}
\end{equation}
where 
\begin{equation}
      T_0^4 = { A \int B_\nu d\nu \over
            \sigma \int B_\nu Q_\nu (h\nu)^{-1} N(h\nu) }
          \label{eqn:t0}
\end{equation}
and $N(h \nu) = 4 \pi a^2 \int R S dt$ is the average number of particles
that are evaporated after a cold grain absorbs a photon of energy 
$h \nu$.  These equations suggest 
a functional form for a successful interpolation formula
\begin{equation}
   T_e(T_{bb},T_*)  = { \alpha(T_*) + \beta(T_*) \left(T_{bb}/1000~\hbox{K}\right)  
     \over 1 + \gamma(T_*) \log_{10}\left(T_{bb}/10~\hbox{K}\right) }
      \label{eqn:fit}
\end{equation}
to the numerical results.  We force the fits to agree with 
Equations~\ref{eqn:limit} and \ref{eqn:t0} as $T_{bb}\rightarrow 0$ and our 
Monte Carlo simulations begin to be noisy.   We made
$\alpha$, $\beta$ and $\gamma$ third order polynomials
of the form $\alpha(T_*)= \sum_{i=0}^3 \alpha_i x^i$
with $x = \log_{10} (T_*/1000~\hbox{K})$.  These typically
fit the $T_e(T_{bb},T_*)$ curves as a function of $T_{bb}$
and $T_*$ with root mean square residuals of less than
$100$~K.  The resulting fits for thermally isolated, stochastically
heated grains (Case 4) are presented in Table~\ref{tab:interp}. 

We computed the effective evaporation temperatures for both black
bodies and Solar metallicity stellar atmosphere models.  
For $T_* \geq 3000$~K we used the \cite{Castelli2004} models and 
for the coldest $T_*= 2600$~K model we used \cite{Gustafsson2008}.  
For cool stars, the harder UV emission may be dominated by hotter 
chromospheric and coronal regions above the photosphere (e.g.,
\citealt{Scalo1980}). These effects could be included by 
using a higher effective temperature that would mimic these
contributions.  This temperature can be estimated for any
spectrum by matching the values of $T_0$ in Equation~\ref{eqn:t0}.

Although we will not make use of the results in the present paper,
we also computed effective evaporation rates for model spectra
of supernovae and quasars.  We used Type~IIP model spectra for
days 98 and 197 from \cite{Dessart2013} and Hillier \& Dessart (2014, 
in prep) which are very similar to those of \cite{Jerkstrand2012}
at the same phase.  This phase corresponds to the early,
post-plateau period when the supernova has faded significantly
and dust formation becomes possible.  For quasars we used the
parametrized spectrum of \cite{Honig2010}.  We again truncate
the spectra at $13.6$~eV.  While the harder X-ray photons from
the quasar will reach the dust formation region, they transfer
little energy to small grains (see \citealt{Voit1991}, 
\citealt{Voit1992}).  The net result is that the effective
evaporation temperatures for these supernova spectra are 
roughly those of a $T_*\simeq 4000$-$6000$~K star and 
the quasar is roughly similar to a $T_*\simeq 25000$~K star.
For supernova there are additional complications for
dust formation in the region where the nuclear
decay energy is being absorbed (e.g., \citealt{Cherchneff2009}),
but we will consider this problem in a later paper.

\begin{figure}
\includegraphics[width=3.5in]{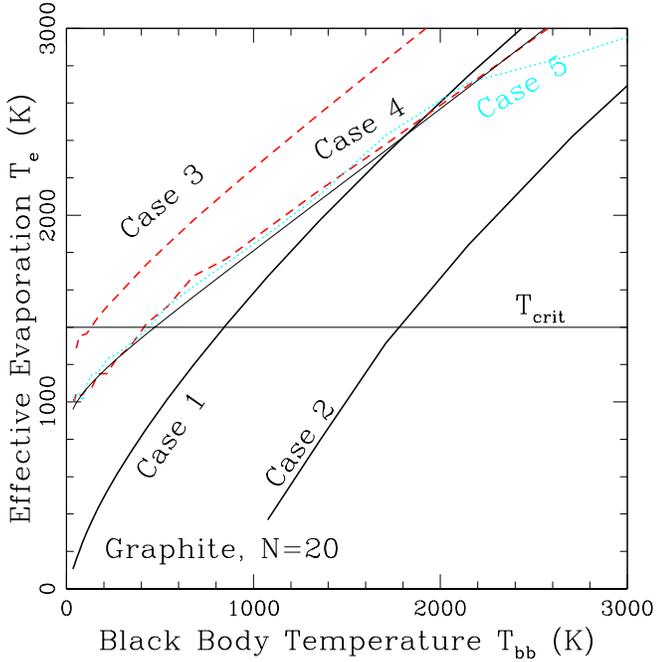}
\caption{
  Effective evaporation temperatures, $T_e$, as a function of the black body
  temperature, $T_{bb}$, for a $T_*=5000$~K black body.  A horizontal
  line shows a typical value of $T_{crit}$ where the grain can start to
  grow once $T_e<T_{crit}$.  The heavy solid lines show
  the two cases using Planck equilibrium temperatures.  Case 1 represents the
  standard model with $S\equiv 1$, while Case 2 shows the effect of thermally
  isolating the grains.  Case 3 shows the effect of stochastically heating
  grains in thermal equilibrium with their carrier gas, which leads to
  a tremendous increase in evaporation rates.  Case 4 then adds the suppression
  factor to thermally isolate the grains, which produces evaporation rates
  intermediate to Cases 1 and 3.  The thin black line shows the analytic
  approximation to this case using the formula in Table~\ref{tab:interp}.
  Finally, Case 5 adds the thermal effects of evaporation and capture, which
  leads to a net reduction in the mean evaporation rate at high temperatures
  compared to Case 4 because evaporation preferentially occurs when the grain is hot.
  }
\label{fig:outline1}
\end{figure}

\begin{figure}
\includegraphics[width=3.5in]{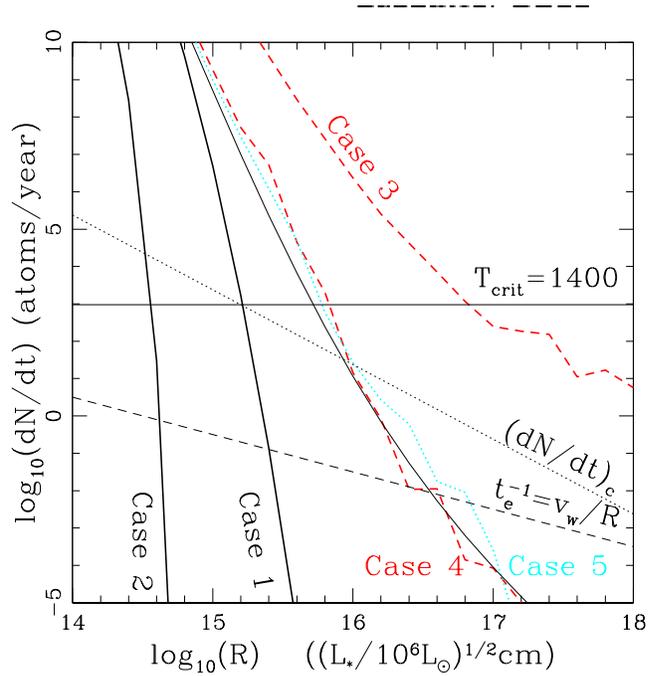}
\caption{
  Evaporation rates as a function of radius for a $L_*=10^6 L_\odot$,
  $T_*=5000$~K black body.  The Cases are the same as in Figure~\ref{fig:outline1}.
  The horizontal line is the collision rate corresponding to $T_{crit}=1400$~K,
  The dotted line labeled $(dN/dt)_c$ shows the collision rate for a wind
  with $\dot{M}\simeq 2 \times 10^5 M_\odot$~year$^{-1}$ and $v_w=100$~km/s,
  and the dashed line labeled $t_e^{-1} = v_w/R$ shows the expansion rate.
  A grain grows if the collision rate is larger than both the evaporation
  rate and the expansion rate.
  Compared to the reference Case 1 (Planck equilibrium),
  Case 2 (thermal isolation) allows dust formation at much smaller radii and Case 3
  (stochastic heating) would allow it only at much larger radii. Case
  4, combining both effects, increases the dust formation radius
  compared to Case 1 by a factor of $\sim 4$.  For the parameter
  range shown, Case 5 is essentially the same as Case 4.
  Case 3 would never form dust for this wind density.
  The radius can be rescaled as $L_*^{1/2}$.
  }
\label{fig:outline2}
\end{figure}

\section{Physical Results}
\label{sec:results}

Figure~\ref{fig:outline1} shows the effective evaporation temperature
$T_e(T_{bb})$ for a $5000$~K black body spectrum and graphitic
grains with $N=20$.   To guide the eye, we include a line with
a constant $T_{crit}=1400$~K.  For a real wind, $T_{crit}$ slowly
drops as a function of $T_{bb}$ because of the density 
dependence in Equation~\ref{eqn:tcrit0}, as we will discuss below.
For the standard, Case 1, Planck radiative 
equilibrium model, this grain could begin to grow when the
black body temperature dropped to $T_{bb} \simeq 860$~K
because the Planck equilibrium temperature of $1400$~K
is significantly higher than $T_{bb}$ due to the large ratio of
$\langle Q_{abs}\rangle/Q_{em}$ for these small grains.
  
The evaporation rate in Case 1 assumes that the grains are in 
a thermodynamic bath of the same temperature.  On Earth this would
be true because the far higher gas densities would keep the grain
temperature in collisional equilibrium with the gas temperature 
on collisional time scales that are short compared to the radiative 
time scales.  In these astrophysical flows, however, the gas collision
rate is slow compared to the photon absorption rate or the radiative cooling
time scale.  Case 2 shows the result of adding the suppression factor
to consider a thermally isolated grain.  In this scenario it becomes
exceedingly difficult to evaporate anything from a small grain
because the enthalpy of the grain
does not approach the critical temperature $k B \simeq 7$~eV 
until $T_g \simeq 2000$~K.  As a result, the grain can begin
growing at $T_{bb} \simeq 1800$~K. 

In Case 3 we modify Case 1 by stochastically heating the grains.
This leads to enormously higher evaporation rates for the same
energy flux ($T_{bb}$).  A $T_*=5000$~K black body emits
75\% (8\%) of its luminosity in photons with energies above 
1~eV (3~eV), and these photons can produce $\Delta T \simeq 600$ ($1100$)~K
temperature spikes for a $N=20$ grain starting at $T_g=0$~K.  As
a result, the effective evaporation temperature remains high
even as $T_{bb}$ drops, and the grain would only be able to
start growing at $T_{bb} \simeq 140$~K if $T_{crit}$ is simply
held fixed at $1400$~K.  Case 3 is particularly unphysical 
because having the grains in thermal equilibrium with the 
gas essentially assumes that the gas temperature fluctuates in 
concert with the grain temperature.  

In reality, we must make the grains both thermally isolated and
stochastically heated, as shown by Case 4.  For this radiation
temperature, the net effect is still an evaporation rate that
is significantly higher than predicted by Case 1.  A thin
black line underlying the curve shows results for the
interpolation formula in Table~\ref{tab:interp} for this
case.  As noted in \S\ref{sec:form}, the interpolation formulae are not
perfect, as seen here from the small differences in curvature.
Nonetheless, where the numerical solution would allow dust
formation at $T_{bb} \simeq 430$~K the interpolation formula
gives $480$~K which corresponds to only a 25\% shift in 
radius.

Finally, Case 4 still lacks the effects of evaporation and
condensation on the grain temperature, and this is shown by
Case 5.  At low temperatures, there is no change from Case 4
because the rate of evaporations and captures is low compared
to the rate of photon absorptions.  At very high temperatures,
the effective evaporation temperature in Case 5 starts to be
lower, because evaporation begins to contribute significantly
to grain cooling during temperature peaks.
This reduces the peak grain temperatures and hence the evaporation
rate over Case 4.  Essentially, Case 5 starts to approach 
the fixed temperature of a boiling liquid. However, the differences are only important
at such high temperatures that there is no scenario in which
the gas densities would be high enough for collision rates to balance
evaporation rates.  
 
Figure~\ref{fig:outline2} shows these results translated into
rates as a function of radius for a $L_*=10^6 L_\odot$ source.  
A horizontal line again
shows the constant collision rate corresponding to $T_{crit}=1400$~K
and a grain will grow once the collision rate is higher than the
evaporation rate.  For fixed $T_{crit}$, Case 1 reaches this collisional rate
at $10^{15.2}$~cm.  Case 2, including the suppression factor
but not stochastic heating, would drop the formation radius
by almost a factor of $5$ to $10^{14.6}$~cm.  Case 3, adding
stochastic heating without the suppression factor, would increase
the formation radius compared to Case 1 by almost a factor of $50$ 
to $10^{16.9}$~cm.
At the largest radii, the slope of the evaporation rate curve
for this case has converged to the $\langle dN/dt \rangle_e \propto r^{-2}$ scaling predicted
in \S1.  Case 4, with both stochastic heating and the suppression
factor, leads to a radius roughly $4$ times larger than in Case 1, at $10^{15.8}$~cm.
Case 5, where we had the heating and cooling from evaporation and
capture, is little different from Case 4 over the
parameter range shown in Figure~\ref{fig:outline2}.   
Given the interpolation formula for $T_e(T_{bb})$, the evaporation 
rate is simply $4\pi a^2 A \exp(-B/T_e(T_{bb}))$ and a thin 
black line shows that this model recreates the numerical
results reasonably well.  

In practice, the collision rate (Equation~\ref{eqn:coll}) and
the required $T_{crit}$ vary with radius because of the 
changing density of a wind.  To illustrate this,
Figure~\ref{fig:outline2} also shows the
collision rate $(dN/dt)_c$ for a wind with 
$\dot{M}\simeq 2 \times 10^{-5} M_\odot$~year$^{-1}$,
$v_w=100$~km/s and $X=0.01$.  This was chosen so that
it passes through the point at $R \simeq 10^{15.2}$~cm
where the Case 1 evaporation rate has $T_{crit}=1400$~K.
The collision rate simply declines with the density
as $(dN/dt)_c \propto R^{-2}$ which 
translates into $T_{crit}=1730$, $1560$, $1430$, $1310$,
and $1210$~K at $R=10^{14}$, $10^{15}$, $10^{16}$, $10^{17}$
and $10^{18}$~cm.  Compared to a fixed $T_{crit}$ (density),
the dust formation
radius becomes slightly smaller for Case 2 and slightly
larger for Cases 4 and 5.  For Case 3, however, the evaporation
rate is above the collision rate at all radii and no dust
would form in this case.  Also shown is the expansion
rate, $t_e^{-1}=v_w/R$, which determines the minimum
mass loss rate for dust formation.  If the density is
so low that the collision rate is below $t_e^{-1}$, the
grain has no further collisions and stops growing.  

\begin{figure}
\includegraphics[width=3.5in]{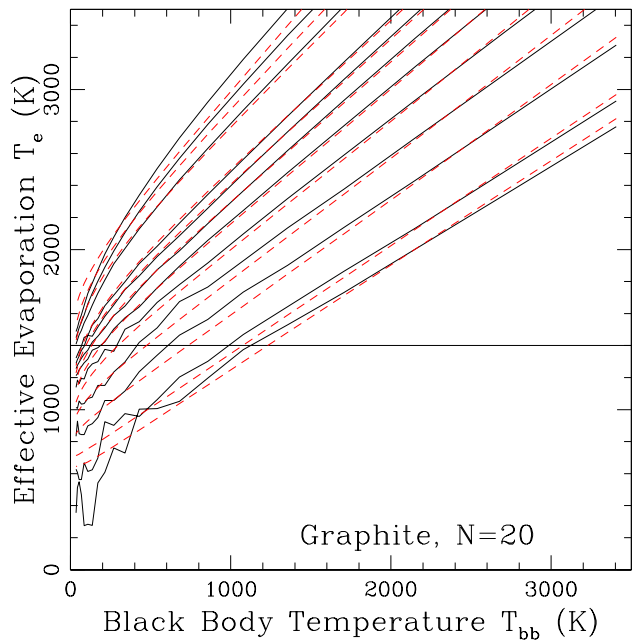}
\caption{
  The Case 4 effective evaporation temperature, $T_e$, as a function of the
  black body temperature, $T_{bb}$, for (from bottom to top) $T_*=2600$,
  $3000$, $4000$, $5000$, $6000$, $7000$, $8000$, $9000$, $10000$, 
  $15000$, $20000$ and $30000$~K black bodies and $N=20$ graphitic
  grains.  The solid lines are the numerical results and the dashed
  lines are the results using the approximation formula for this case
  from Table~\ref{tab:interp}.  The root mean square difference 
  between the two is $80$~K, which is partly due to the order of
  the approximation and partly due to the noise in the Monte Carlo
  results at low $T_{bb}$ and $T_*$.
  }
\label{fig:compare}
\end{figure}

\begin{figure}
\includegraphics[width=3.5in]{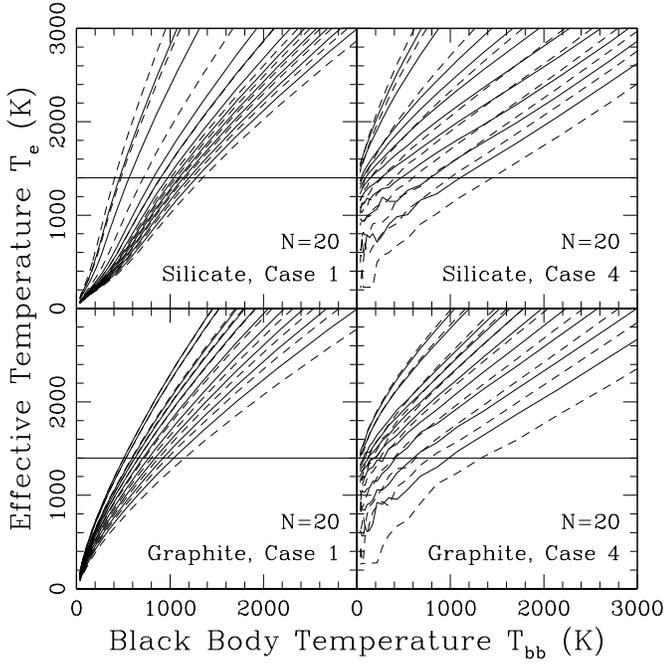}
\caption{
  The effective evaporation temperature, $T_e$ as a function of $T_{bb}$
  for $N=20$ graphitic (top) and silicate (bottom) grains with
  radiation temperatures (from bottom to top) of  $T_*=2600$,
  $3000$, $4000$, $5000$, $6000$, $7000$, $8000$, $9000$, $10000$,
  $15000$, $20000$ and $30000$~K.  The horizontal line at
  $T_{crit}=1400$~K corresponds to a typical collision rate. 
  The solid lines are for black
  bodies and the dashed lines are for stellar atmosphere models. 
  The left panels show the Planck equilibrium results (Case 1)
  and the right panels show the results for stochastically
  heated, thermally isolated grains (Case 4).   
  }
\label{fig:n20}
\end{figure}

\begin{figure}
\includegraphics[width=3.5in]{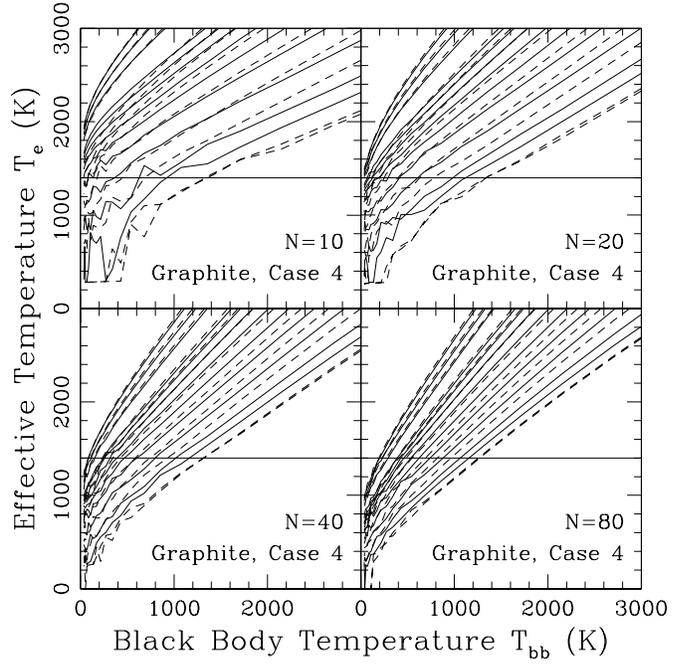}
\caption{
  The effective evaporation temperature, $T_e$ as a function of $T_{bb}$
  for stochastically heated, thermally isolated (Case 4) $N=10$ (top left), 
  $20$ (top right), $40$ (lower left) and $80$ (lower right) graphitic grains   
  The range of radiation temperatures
  is the same as in Figure~\ref{fig:n20} for both black bodies
  (solid) and stellar atmospheres (dashed).  The horizontal line at
  $T_{crit}=1400$~K corresponds to a typical collision rate.
  }
\label{fig:ngraphite}
\end{figure}

\begin{figure}
\includegraphics[width=3.5in]{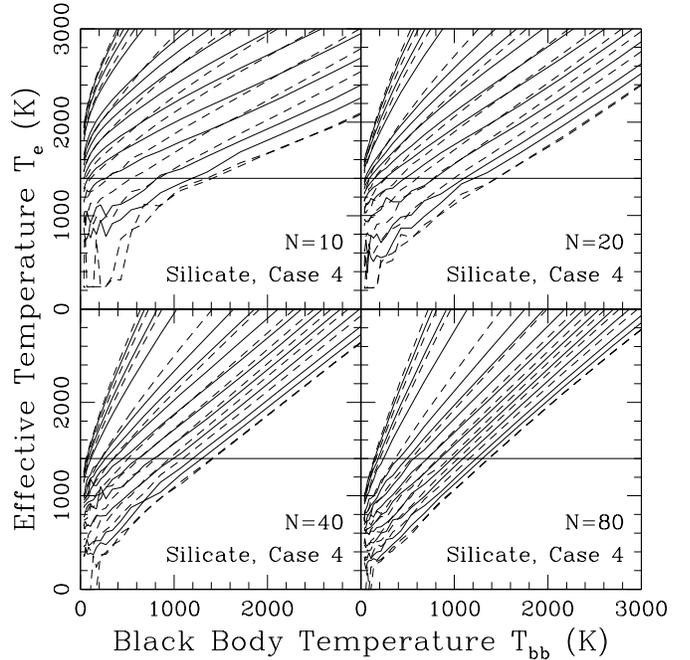}
\caption{
  As in Figure~\ref{fig:ngraphite}, but for silicate grains.
  }
\label{fig:nsilicate}
\end{figure}

We will now only focus on the differences between Case 1, representing
the standard Planck radiative equilibrium model, and Case 4.  We
do not discuss Cases 2 and 3 further because they are not physical, 
and Case 5 differs from Case 4 only under physical conditions that
do not seem to be relevant for dust formation.  Figure~\ref{fig:compare}
shows the full suite of graphitic Case 4 $N=20$ black body
models models and the interpolated model from Table~\ref{tab:interp}
for radiation temperatures from $T_*=2600$~K to $30000$~K.  The
overall rms residual is $80$~K, partly due to systematic mismatches
and partly due to noise for the low $T_{bb}$ and $T_*$ cases.  Most
of the differences can be thought of as small mismatches ($\sim 10\%$)
in the meaning of $T_*$ between the numerical models and the 
approximations.  Since it is unlikely that these approximations will 
ever be used in situations where the temperature is that well
known, and because of the additional uncertainties in the input
dust physics, there seemed no need to develop still more complicated
approximations that would significantly reduce the residuals. 
The results for the other models are similar, so we will not
show further comparisons of the numerical results and the
approximations.

Figure~\ref{fig:n20} shows the Case 1 and Case 4 results for
both graphitic and silicate $N=20$ grains, both black bodies
and model atmospheres, and as a function of radiation 
temperature $T_*$.  A constant $T_{crit}=1400$~K line again
provides a comparison.  Without stochastic heating, the
effective evaporation temperature does depend on $T_*$ in
the sense that $T_e$ increases with $T_*$ at all fluxes.  This
is simply driven by the rise in $\langle Q_{abs} \rangle$
with $T_*$.  The trends are relatively smooth except for
a faster rise for silicate grains and high $T_*$ created
by the jump in $Q_{abs}$ near $\lambda \simeq 0.2\mu$m.
At low temperatures, stellar atmosphere models have lower
$T_e$ for fixed $T_*$ because of strong atmospheric 
absorption at shorter wavelengths for cooler stars.  The
effect becomes smaller for hotter stars because the 
spectral breaks become weaker. 

For the coolest stars, the additional physical effects 
essentially cancel to leave little difference between
Case 1 and Case 4.  For the $T_*=2600$~K stellar
atmospheres, the silicate grains have $T_e=1400$~K
at $T_{bb} \simeq 1400$~K in both cases, while for
the graphitic grains the temperature rises modestly
from $T_{bb} \simeq 1200$ to $1400$~K, which means
the grain would actually start to grow sooner than
in Case 1.  However, even for moderately higher radiation 
temperatures, Case 4 requires substantially lower
radiation fluxes ($T_{bb}$) for the same evaporation
rate.  We have already discussed the $T_*=5000$~K case 
in detail, and for higher radiation temperatures the
differences become still more extreme.
Figures~\ref{fig:ngraphite} and ~\ref{fig:nsilicate}
show the changes in $T_e$ with grain size for
$N=10$, $20$, $40$ and $80$.  Not surprisingly, the
effects of stochastic heating are stronger for 
smaller grains and weaker for larger grains. While
we do not show the comparisons, the Case 4 results
increasingly resemble the Case 1 results as $N$
increases, although even for $N=80$ there are 
significant differences and the differences become larger
for higher radiation temperatures.  

\begin{figure}
  \includegraphics[width=3.4in]{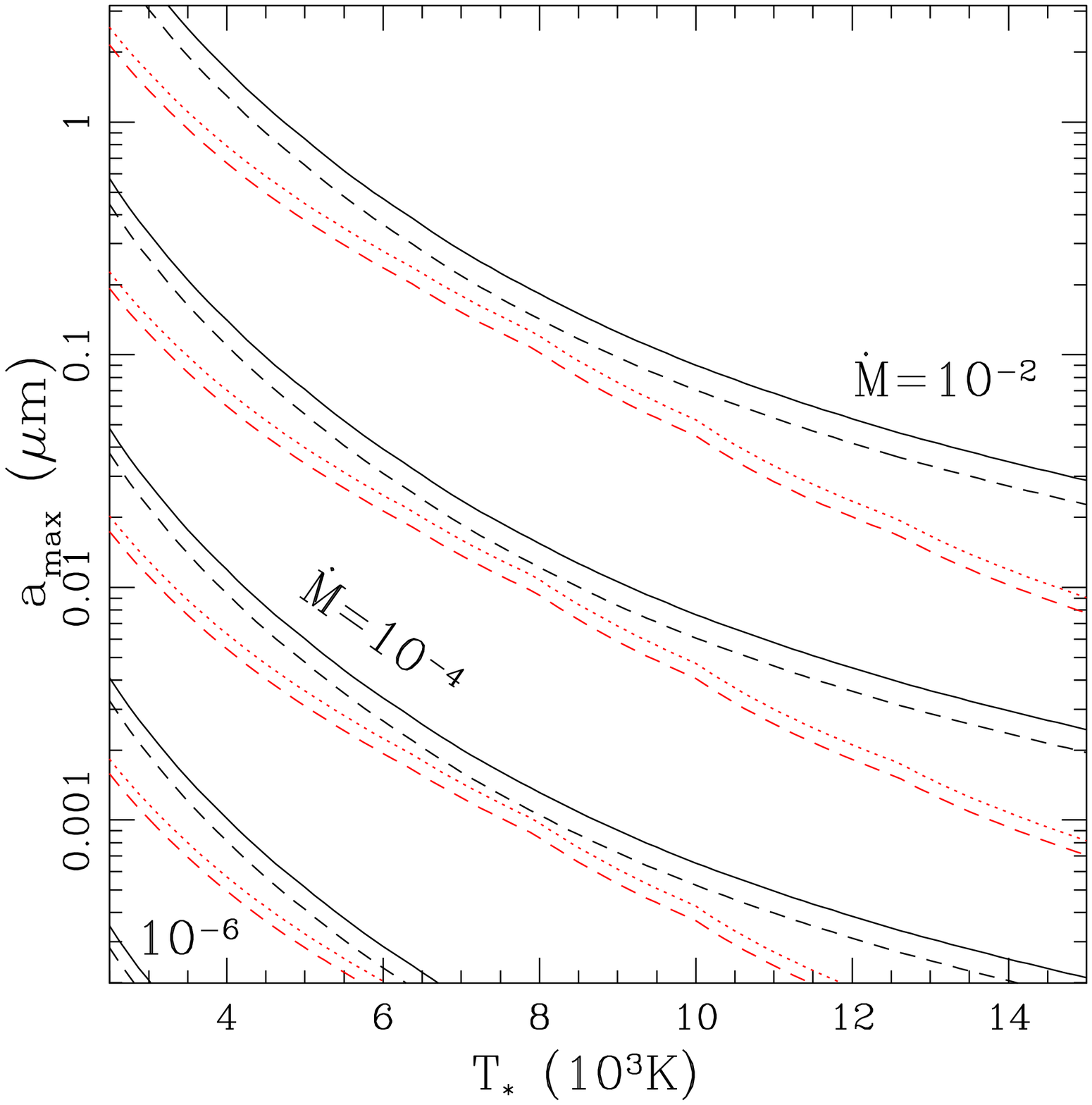}
  \caption{
    Final grain sizes $a_{max}$ assuming Planck equilibrium
    temperatures (Case 1) and black body spectra as a function 
    of stellar temperature.   The
    solid black (dotted red) curves are for graphite (silicate)
    grains and $L_*=10^4 L_\odot$.  The adjacent but slightly
    lower black (red) dashed curves are for graphite (silicate)
    and $L_*=10^6 L_\odot$.  The
    mass loss rates, from largest to smallest final sizes,
    are $\dot{M}=10^{-2}$, $10^{-3}$, $10^{-4}$, $10^{-5}$
    and $10^{-6}M_\odot$/year with the dependencies on
    other variables discussed in the text. The nominal
    case is $N=20$ with $\alpha=1$, $\beta=1$,
    $X=0.005$, $v_c=1$~km/s and $M_*=10 M_\odot$.
    }
  \label{fig:amax1}
  \includegraphics[width=3.4in]{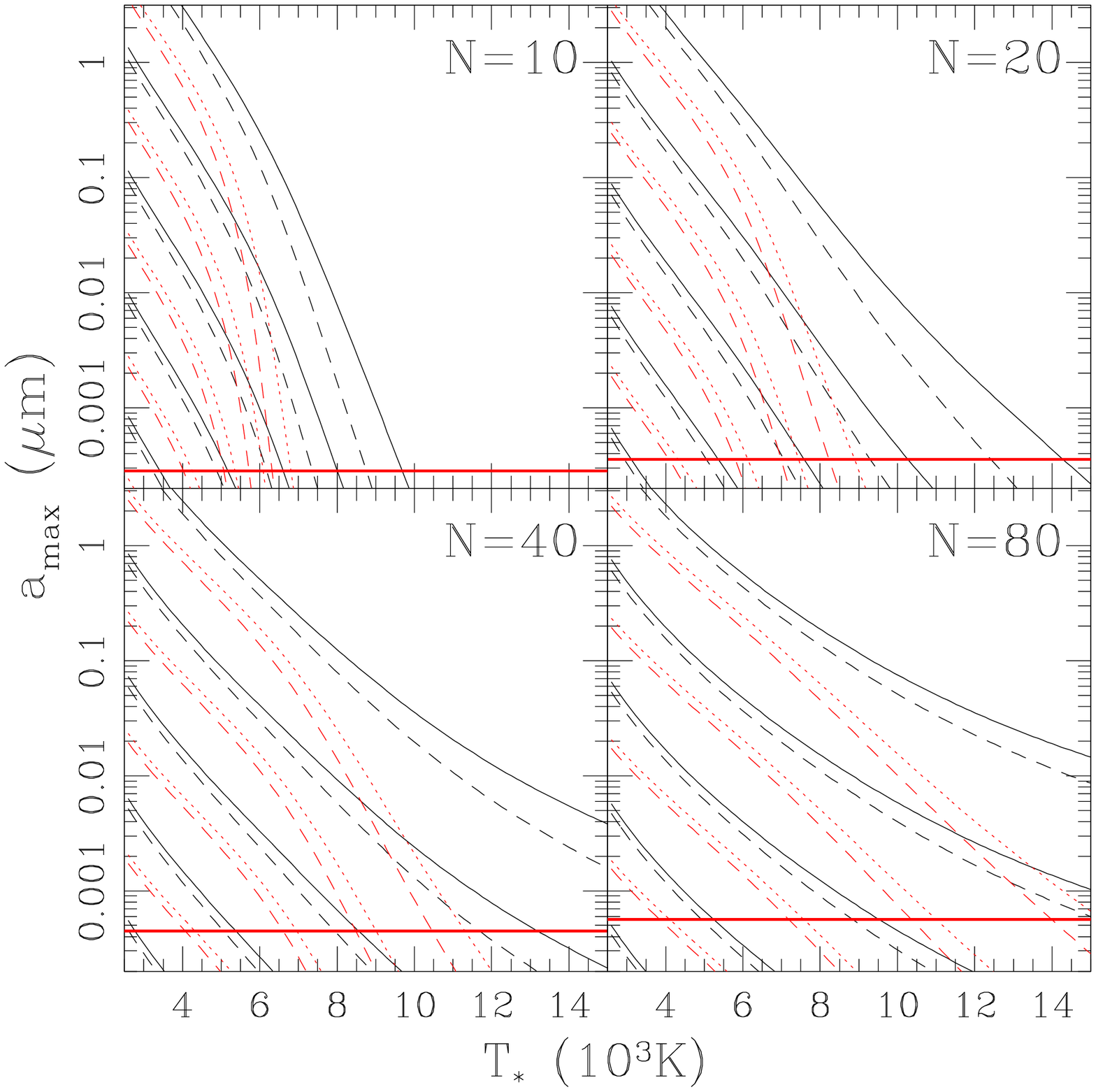}
  \caption{
    Final grain sizes $a_{max}$ assuming stochastically heated,
    thermally isolated grains (Case 4) and stellar atmospheres
    as a function
    of stellar temperature for the $N=10$ (top left), $20$ (top right)
    $40$ (lower left) and 80 (lower right) cases.  The
    solid black (dotted red) curves are for graphite (silicate)
    grains and $L_*=10^4 L_\odot$.  The adjacent but slightly
    lower black (red) dashed curves are for graphite (silicate)
    and $L_*=10^6 L_\odot$.  The mass loss rates and parameters
    other than $N$ are as in Figure~\protect\ref{fig:amax1}.
    The heavy horizontal line shows the size of an $N$
    particle grain. 
    }
  \label{fig:amax2}
\end{figure}

\begin{figure}
\includegraphics[width=3.5in]{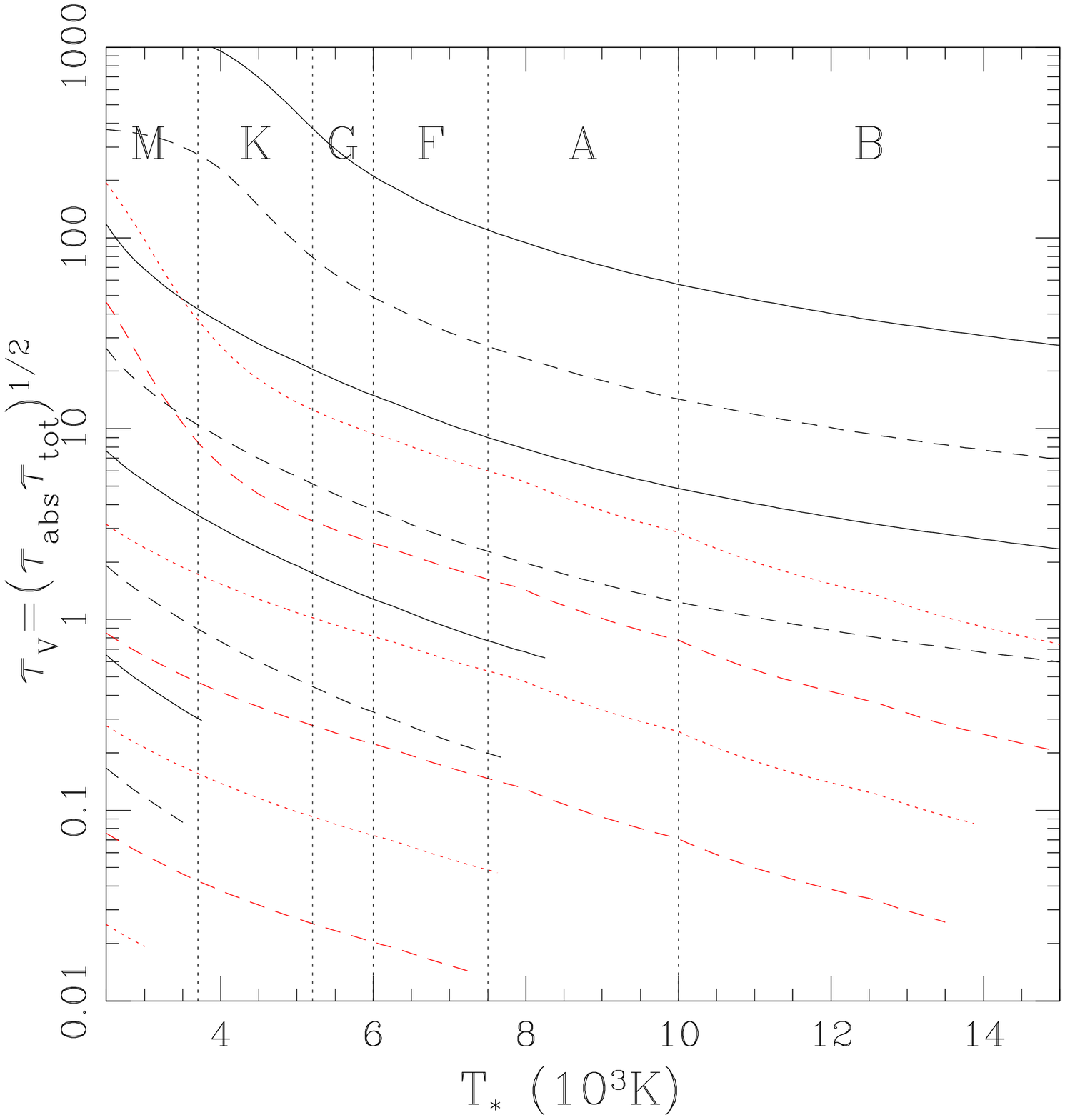}
\caption{
  Wind optical depths at V-band assuming Planck equilibrium (Case 1).
  The solid black (dotted red) curves are for graphite (silicate)
  grains and $L_*=10^4 L_\odot$.  The adjacent but 
  lower black (red) dashed curves are for graphite (silicate)
  and $L_*=10^6 L_\odot$.  The
  mass loss rates, from largest to smallest optical depth,
  are $\dot{M}=10^{-3}$, $10^{-4}$, $10^{-5}$
  and $10^{-6}M_\odot$/year with the dependencies on
  other variables discussed in the text. The nominal 
  case is $N=20$ and the curves are truncated when
  $a_{max}$ corresponds to $N=20$.
  }
\label{fig:tau1}
\includegraphics[width=3.5in]{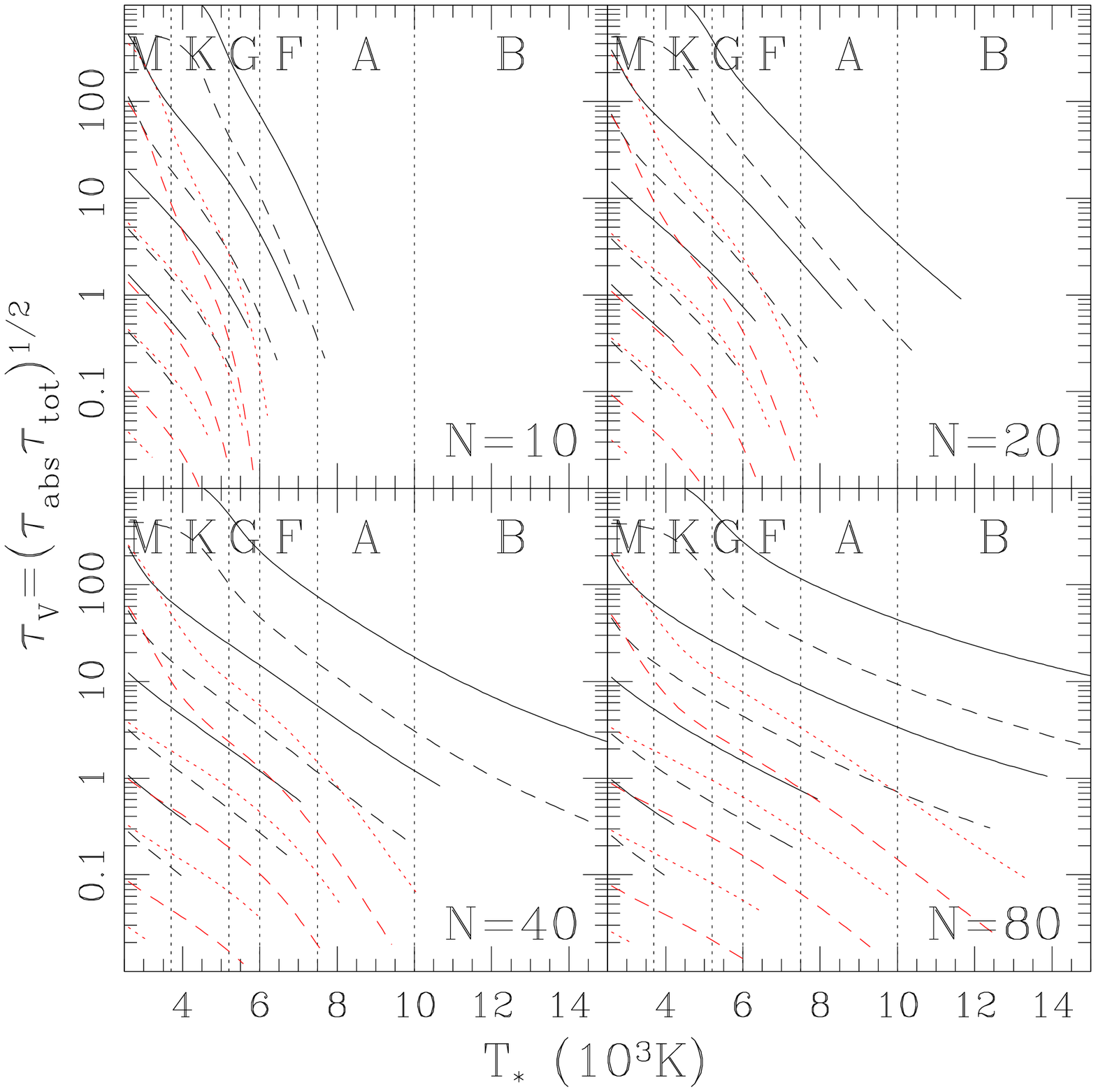}
\caption{
  Wind optical depths at V-band assuming thermally isolated,
  stochastically heated grains (Case 4) for $N=10$, $20$,
  $40$ and $80$.
  The solid black (dotted red) curves are for graphite (silicate)
  grains and $L_*=10^4 L_\odot$.  The adjacent but 
  lower black (red) dashed curves are for graphite (silicate)
  and $L_*=10^6 L_\odot$.  The
  mass loss rates, from largest to smallest optical depth, 
  are $\dot{M}=10^{-3}$, $10^{-4}$, $10^{-5}$
  and $10^{-6}M_\odot$/year with the dependencies on
  other variables discussed in the text. The curves
  are terminated when $a_{max}$ corresponds to $N$
  particles. 
  }
\label{fig:tau2}
\end{figure}

\section{Dust Formation In Winds}
\label{sec:winds}

For a steady wind we can integrate Equation~\ref{eqn:coll} from 
the radius $R_f$ where the growth rate begins to exceed
the evaporation rate to determine the typical final grain size.
If we scale the wind velocity by the escape velocity 
(Equation~\ref{eqn:vesc}) then  
\begin{equation}
    a_{max} = { \alpha X \dot{M} v_c \over 32 \pi G M_* \rho_b \beta^2}
            \left( { T_{bb} \over T_* } \right)^2
        \simeq 0.16 { \alpha X_2 \dot{M}_4  v_{c1} \over \beta^2 M_{*10} \rho_{b3} }
            \left( { T_{bb} \over T_* } \right)^2~\mu\hbox{m}
       \label{eqn:amax}
\end{equation}
where the formation radius is characterized by the black body 
temperature $T_{bb}$ at which the grains can begin to grow.
In order to be considered a dust particle, this final size must
contain some minimum number of particles, $N_{min}=10N_{min10}$, which 
implies a minimum mass loss rate
\begin{eqnarray}
   \dot{M} &> &8 \left( 6 \pi^2 N_{min} m_0 \rho_b^2 \right)^{1/3} 
         { G M_* \beta^2 \over \alpha X v_c } \left( { T_* \over T_{bb}} \right)^2 \\
     &\gtorder &6 \times 10^{-8} { N_{min10}^{1/3} M_{*10}\beta^2  \over \alpha v_{c1} X_2 }
            \left( { T_* \over T_{bb} } \right)^2~M_\odot/\hbox{year},
         \nonumber
\end{eqnarray}
below which the growth of the grain is prevented by the kinetics
of the collisional process.  Note that to lowest order, the 
final dust properties are independent of the stellar luminosity
and are largely controlled by $\dot{M}$ and the 
temperature ratio $T_{bb}/T_*$.  Since the black body temperature
$T_{bb}$ at which dust formation begins also is a function of
$T_*$, the dependence of the dust properties on $T_*$ is even
stronger than the explicit quadratic.

Equation~\ref{eqn:amax} does not represent a true model of a 
dust accelerated wind because we are primarily
interested in dust formation in stellar eruptions, supernovae or
AGN where the dust is not responsible for accelerating the gas
to produce the outflow.  Despite ignoring the problem of self-consistently
accelerating a dusty wind, the simple model based on Equation~\ref{eqn:amax}
produces quite reasonable results.  For example, if we simply
apply Equation~\ref{eqn:amax} with $X=0.0046$, $M_*= 1M_\odot$
and $L_* = 2\times 10^4 L_\odot$ to match the $\epsilon_c/\epsilon_0=1.6$
case of \cite{Gail1985}, we find $a_{max}=0.006$, $0.15$ and $0.84\mu$m,
compared to $0.013$, $0.031$ and $0.13\mu$m in the full calculations
for $\dot{M}=10^{-6}$, $2\times 10^{-5}$ and $2\times 10^{-4}M_\odot$~year.
This also assumed $\beta=1$, $v_c=1$~km/s, $T_*=2600$~K and Planck
equilibrium (Case 1).  The agreement is relatively good, and the
differences are likely due to ignoring the effect of the dust on
the radiation field. Once the dust has an appreciable optical
depth, Equation~\ref{eqn:equil} is incomplete because as the
newly formed dust begins to trap some of the radiation, the temperatures
rise and the dust formation radius is driven outwards
leading to smaller final grain sizes.  
In experiments with DUSTY (\citealt{Ivezic1997}, \citealt{Ivezic1999}),
we found that the formation radius increases roughly as $R_f \propto (1+\tau_V)^{0.2}$
due to this effect.  For $\tau_V\simeq 100$, like the $\dot{M}=2\times 10^{-4}M_\odot$~year
model from \cite{Gail1985}, this would lead us to overestmate the grain size
by a factor of $2$-$3$, accounting for much of the difference in the estimates.
This effect is not critical to the discussion which follows, because it only
becomes important once dust forms with an appreciable optical
depth, and we are mainly concerned with the question of whether
dust forms at all.  

Figure~\ref{fig:amax1} shows the dependence of $a_{max}$ on mass
loss rates and stellar temperature using Planck equilibrium
temperatures (Case 1) and black bodies. Figure~\ref{fig:amax2} shows the
results for stochastically heated, thermally isolated grains
(Case 4) and stellar atmospheres. For the Planck equilibrium
case we only show $N=20$ because the other cases will be almost
identical.  These results use $X\equiv 0.005$, $\beta=1$, 
$\alpha \equiv 1$, $v_c \equiv 1$~km/s and  $M_* \equiv 10 M_\odot$.
To first order, the final grain size simply scales as in 
Equation~\ref{eqn:amax} but there is a secondary, logarithmic
effect from the dependence of $T_{crit}$ on the variable
combination $X \dot{M} v_c / \beta M_*^{1/2} L_*^{3/4}$ 
(Equations~\ref{eqn:tcrit1} and \ref{eqn:tcrit2}).  We illustrate
these effects by showing the results for $L_*=10^4$ and $10^6 L_\odot$.  

If we compare Figure~\ref{fig:amax1} and \ref{fig:amax2}, we see that many LBVs and
all yellow supergiants in their hotter states could form 
dust if we consider Case 1 (Figure~\ref{fig:amax1}).  However,
particularly for the oxygen rich, silicate chemistry usually
associated with the winds of massive stars, they cannot do
so if we consider Case 4 (Figure~\ref{fig:amax2}).
There is some ambiguity as to the appropriate $N$, but
essentially any of the four cases shown will reproduce the
observed behavior.  As hot O/B (LBV) or A (yellow supergiants)
stars, it is impossible for these stars to form dust even at
mass loss rates near $\dot{M}\sim 10^{-4}M_\odot$/year.
When LBVs enter their eruptive states, the wind becomes (non-dust)
optically thick and the stellar photosphere is shielded by a
pseudo-photosphere created by the wind leading to the apparent
temperature of an F star ($T_* \sim 7500$~K, \citealt{Davidson1987})
and at this radiation temperature it becomes possible to form
dust (\citealt{Kochanek2011}).  Similarly, in their cooler
states, the yellow hypergiants appear as cooler G/F stars  
and also enter the regime where dust formation is allowed.
Particularly for the yellow hypergiants it is very difficult
to understand the lack of dust formation in their hotter phases
without this mechanism.

We can also estimate the resulting optical depth assuming that 
at every radius all the condensible material is in grains 
of size $a= a_{max}(1-R_f/R)$ found by integrating 
Equation~\ref{eqn:coll} starting at the formation radius
$R_f$ where growth begins.  This is slightly inconsistent
because it assumes monomer growth and a constant collision
velocity -- for example, if all grains are in grains of 
size $a$ than growth is really be coagulation with a
geometric cross section of $4\pi a^2$ rather than $\pi a^2$.
Similarly, sticking probabilities and collision velocities
will also change.  However, the biggest uncertainty is how
many grains are initially formed to begin growing, which
we encapsulate as the fraction $f$ of the condensible 
material that is ultimately incorporated into grains
(see, e.g., the discussion in \citealt{Ferrarotti2006}). 
Recent detailed simulations by \cite{Nanni2013} and
\cite{Ventura2014}  find condensed fractions of order 
$f \sim 0.1$ to $0.5$.  The dust represents mass fraction 
$X f$ of the total wind mass.  Observationally, this
is roughly $ X f \sim 1/200$ to $1/400$ for Galactic
sources (e.g., \citealt{Knapp1985}).  Since we have scaled
our results to $X=0.005=1/200$ typical values of $f$
should be in the range $0.5 \ltorder f \ltorder 1$.

Under these assumptions, the opacity is just 
$ \kappa = 3 f X Q/4\rho_b a$, so the optical depth is 
\begin{equation}
      \tau = { 3 f v_w \over \alpha v_c} \int_0^{a_{max}} Q(a) da/a
     \label{eqn:tau}
\end{equation} 
after converting the radial integral into one over the grain size.  
When the grains are small, $Q/a$ is nearly constant, so the
optical depth simply grows $\propto a_{max}$.  It rises more steeply
for $0.1 < a_{max} < 1\mu$m and then flattens for still larger
grains.  In any dust forming flow there should always be a
strong correlation of optical depth with maximum grain size.
Note that since $a_{max} \propto v_w^{-2}$, the optical depth
does decrease with increasing wind velocity despite the leading
factor of $v_w$ in Equation~\ref{eqn:tau}.  We estimate the effective
absorption optical depth as $\tau = (\tau_{abs}\tau_{tot})^{1/2}$
where $\tau_{tot}=\tau_{abs}+\tau_{sca}$ combines the absorption
and scattering optical depths.
  
Figure~\ref{fig:tau1} shows the resulting visual optical depths
assuming Planck equilibrium (Case 1) as a function of stellar 
temperature, luminosity and mass loss rate. We assume $f=1$
and truncate the results when $a_{max}$ corresponds to a grain
with $N<20$ particles.   For a given mass loss rate, optical depths 
are highest for colder, lower luminosity stars and graphitic dusts.  
Figure~\ref{fig:tau1} shows mass loss rates of 
$\dot{M}=10^{-6}$, $10^{-5}$, $10^{-4}$ and $10^{-3}M_\odot$/year 
covering the range from stellar winds to the mass loss rates
invoked for SN impostors or to 
explain the superluminous Type~IIn SNe (e.g., \citealt{Ofek2014}).
Without the effects of stochastic heating, most high mass loss
transients would be 
completely opaqued as soon as dust formation becomes feasible.  
Note, however, that other than hotter stars requiring higher
minimum mass loss rates, there is nothing special about the
stellar temperature.

Figure~\ref{fig:tau2} shows the results for thermally isolated,
stochastically heated grains (Case 4) as a function of the limiting 
particle size $N$. Particularly
for the smaller $N$ cases, the expected optical depth drops
sharply as the stellar temperatures rise, and dust formation
simply does not occur for hotter stars and the mass loss rates
of even very dense winds.  Silicate dusts in particular show
an abrupt cutoff at $T_* \sim 7000$~K.  At very low temperatures,
the optical depths are slightly enhanced because the thermal
isolation allows dust formation to begin earlier, as discussed
in \S\ref{sec:results}, although this neglects the physics of accelerating true dust driven
winds.        

\section{Discussion}
\label{sec:discussion}

Dust formation is a balance between growth by collisions
and evaporation. The evaporation rate is set by the grain
temperature and the grain temperature is almost always controlled 
by radiative heating and cooling rather than gas collisions. For small grains, this
temperature is stochastic and driven by the flux of soft
UV photons, just as in the interstellar medium.  Thus,
dust formation is far more dependent on the spectrum
of the radiation field than would be predicted simply
using equilibrium temperatures.  While Planck factors
and stellar wind speeds disfavor dust formation around
hot stars, all stars with $T_* \ltorder 15000$~K and 
$\dot{M} \gtorder 10^{-5}M_\odot$/year should have 
winds with significant visual optical depths $\tau_V \gtorder 0.1$ 
if there were no additional physics.

Stochastic heating provides that extra physics because
the abundance of soft UV photons grows (initially) 
exponentially with stellar temperature.  For a cold
AGB star, stochastic heating is almost irrelevant,
but is is already important for G/F stars, A stars
can only form dust if $\dot{M} \gtorder 10^{-4}M_\odot$/year,
and B stars cannot form dust.  These limits are stronger
for silicate dusts because they have lower evaporation
temperatures.  The primary result from this paper is the
set of approximations presented in Table~\ref{tab:interp}
for the effective evaporation temperatures (Equation~\ref{eqn:tevap}) 
of small grains as a function grain size $N$, radiative
flux as characterized by $T_{bb}$ (Equation~\ref{eqn:equil})
and the radiation temperature $T_*$.   Combined with the
critical temperatures defined by Equations~\ref{eqn:tcrit1}
and $\ref{eqn:tcrit2}$, these can be used to better estimate
whether dust formation can occur and the radius or time at
which it commences.  Note that the apparent temperature of
the newly forming dust will be high because most of the
emission occurs during the temperature spikes rather than
the cooler phases.

The role of stochastic heating in dust formation
naturally explains the episodic dust formation
of LBV stars and OH/IR stars (\citealt{Kochanek2011}).  
At some phases, these
stars appear as hotter B or A stars, respectively,
and cannot form dust despite their high mass loss 
rates.  At some phases they appear as cooler G/F 
stars and form large quantities of dust.  For OH/IR
stars this involves only modest changes in luminosity
and mass loss rates.  In the (great) eruptions of LBVs,
there is a large change in luminosity and mass loss
rates, but the higher mass loss rate matters more
because it allows a lower temperature pseudo-photosphere
to form in the wind (\citealt{Davidson1987}) than 
because the higher wind density increases collisional
growth rates. 

As discussed by \cite{Johnson1993} there are
ultimately problems with simply extrapolating
these approaches to very small $N$.  This has
been discussed most extensively for polycyclic
aromatic hydrocarbons (PAHs) in the interstellar 
medium (e.g., \citealt{Omont1986}).   
The physics of evaporation is roughly
correct for all $N$ since is based on a quantum mechanical
view of vibrational modes, but the binding
energies of carbon atoms in a small molecule
are stronger than for bulk graphite.  The
real break down is in the extrapolation of the
radiation absorption cross sections to small $N$.
Grains absorb at ``all'' wavelengths, while simple
molecules absorb only at discrete wavelengths.  
For example, the electronic transition absorption cross sections 
of PAHs and carbon clusters calculated by \cite{Malloci2007}
begin to break up into discrete bands for $N<10$,
with peaks that are comparable to extrapolating
the bulk $Q$ but now with significant gaps.  
Effectively, as $N$ becomes smaller, $Q$ will 
decline faster than $a$, allowing the formation
of molecules but preventing
their growth into macroscopic grains.

Our ultimate goal for these calculations is to 
examine dust formation in other environments where the 
radiation field is very different from that of cool (AGB) 
stars.  For example, there are models which propose that
dust forms in outflows from quasar accretion disk 
(e.g., \citealt{Elvis2002} \citealt{Czerny2011}).
But the UV radiation field of a quasar has the equivalent
stochastic heating power of a $T_*=25000$~K star even after
cutting off all radiation above $13.6$~eV, and this probably
make dust formation in such outflows impossible.  This
is already implicit in models using these effects to
destroy PAHs in dust near quasars in order to explain
the lack of PAH features in quasar dust spectra (see,
e.g., \citealt{Voit1991}, \citealt{Voit1992}).  

Supernovae
have softer spectra, but their typical radiation temperatures
correspond to $T_*\simeq 4000$-$6000$~K, a regime where small changes
in radiation temperature have a major impact on dust formation.  This
may be related to the contradictory evidence about whether
supernovae are a significant source of dust -- many show no
evidence for significant dust formation, while some clearly
produce large quantities of dust (see the review in \citealt{Gall2011}).    
The stellar transients called supernova ``impostors'' (see
the recent studies by \cite{Smith2011} and \cite{Kochanek2012})
and the pre-supernova outbursts associated with some 
Type~IIn supernovae or needed to provide the massive shells 
proposed to power superluminous Type~IIn supernovae 
(see, e.g., \citealt{Ofek2014}) also have 
radiation temperatures in this regime, so dust formation
should be a powerful probe of the amount of ejected mass. 

\section*{Acknowledgments}
I would like to thank T. Miller, T. Thompson and B. Wyslouzil for valuable discussions.

\vfill\eject

\onecolumn
\begin{deluxetable}{llllrrrr}
\scriptsize
\tablecaption{Fitting Formulas for the Effective Evaporation Temperature}
\tablewidth{0pt}
\tablehead{
  \multicolumn{1}{c}{Spectrum} &
  \multicolumn{1}{c}{Dust} &
  \multicolumn{1}{c}{$N$} &
  \multicolumn{1}{c}{Term} &
  \multicolumn{4}{c}{Coefficients ($\alpha_i$, $\beta_i$, $\gamma_i$)}
  }
\startdata
BB &Gra &10 &$\alpha$ &$390.842$ &$1018.267$ &$-0.541$ &$-0.406$ \\
      &      &   &$\beta$  &$ 1.386$ &$-2.373$ &$ 1.182$ &$ 0.016$ \\
      &      &   &$\gamma$ &$6.3f$ &$-0.279$ &$ 0.869$ &$-0.639$ \\
BB &Sil &10 &$\alpha$ &$511.015$ &$869.375$ &$-0.390$ &$-0.126$ \\
      &      &   &$\beta$  &$ 1.158$ &$-2.059$ &$ 0.919$ &$ 0.082$ \\
      &      &   &$\gamma$ &$6.3f$ &$-0.193$ &$ 0.613$ &$-0.376$ \\
BB &Gra &20 &$\alpha$ &$315.666$ &$804.721$ &$-0.569$ &$-0.249$ \\
      &      &   &$\beta$  &$ 1.718$ &$-2.566$ &$ 0.962$ &$ 0.199$ \\
      &      &   &$\gamma$ &$6.3f$ &$-0.329$ &$ 0.939$ &$-0.642$ \\
BB &Sil &20 &$\alpha$ &$319.523$ &$816.994$ &$-0.425$ &$-0.171$ \\
      &      &   &$\beta$  &$ 1.686$ &$-2.443$ &$ 0.434$ &$ 0.550$ \\
      &      &   &$\gamma$ &$6.3f$ &$-0.265$ &$ 0.710$ &$-0.383$ \\
BB &Gra &40 &$\alpha$ &$218.575$ &$665.769$ &$-0.453$ &$-0.241$ \\
      &      &   &$\beta$  &$ 2.006$ &$-2.672$ &$ 0.654$ &$ 0.351$ \\
      &      &   &$\gamma$ &$6.3f$ &$-0.357$ &$ 0.959$ &$-0.605$ \\
BB &Sil &40 &$\alpha$ &$119.764$ &$784.284$ &$-0.354$ &$-0.148$ \\
      &      &   &$\beta$  &$ 2.260$ &$-2.918$ &$ 0.006$ &$ 0.952$ \\
      &      &   &$\gamma$ &$6.3f$ &$-0.350$ &$ 0.851$ &$-0.428$ \\
BB &Gra &80 &$\alpha$ &$150.741$ &$570.282$ &$-0.471$ &$-0.195$ \\
      &      &   &$\beta$  &$ 2.185$ &$-2.646$ &$ 0.445$ &$ 0.447$ \\
      &      &   &$\gamma$ &$6.3f$ &$-0.368$ &$ 0.936$ &$-0.554$ \\
BB &Sil &80 &$\alpha$ &$-34.368$ &$752.842$ &$-0.400$ &$-0.111$ \\
      &      &   &$\beta$  &$ 2.658$ &$-3.106$ &$-0.326$ &$ 1.281$ \\
      &      &   &$\gamma$ &$6.3f$ &$-0.378$ &$ 0.818$ &$-0.340$ \\
Star &Gra &10 &$\alpha$ &$-143.151$ &$1397.619$ &$-0.417$ &$-0.159$ \\
      &      &   &$\beta$  &$ 1.773$ &$-2.638$ &$ 0.555$ &$ 0.459$ \\
      &      &   &$\gamma$ &$6.3f$ &$-0.449$ &$ 1.296$ &$-0.938$ \\
Star &Sil &10 &$\alpha$ &$-110.286$ &$1360.755$ &$-0.408$ &$-0.168$ \\
      &      &   &$\beta$  &$ 2.015$ &$-2.901$ &$ 0.262$ &$ 0.698$ \\
      &      &   &$\gamma$ &$6.3f$ &$-0.431$ &$ 1.095$ &$-0.635$ \\
Star &Gra &20 &$\alpha$ &$-85.472$ &$1082.070$ &$-0.484$ &$-0.138$ \\
      &      &   &$\beta$  &$ 2.102$ &$-2.947$ &$ 0.255$ &$ 0.744$ \\
      &      &   &$\gamma$ &$6.3f$ &$-0.488$ &$ 1.367$ &$-0.941$ \\
Star &Sil &20 &$\alpha$ &$-155.750$ &$1160.095$ &$-0.518$ &$-0.128$ \\
      &      &   &$\beta$  &$ 2.494$ &$-3.345$ &$-0.152$ &$ 1.123$ \\
      &      &   &$\gamma$ &$6.3f$ &$-0.488$ &$ 1.203$ &$-0.672$ \\
Star &Gra &40 &$\alpha$ &$-44.395$ &$861.750$ &$-0.420$ &$-0.124$ \\
      &      &   &$\beta$  &$ 2.333$ &$-3.030$ &$ 0.018$ &$ 0.865$ \\
      &      &   &$\gamma$ &$6.3f$ &$-0.498$ &$ 1.335$ &$-0.863$ \\
Star &Sil &40 &$\alpha$ &$-330.717$ &$1155.925$ &$-0.362$ &$-0.029$ \\
      &      &   &$\beta$  &$ 3.064$ &$-3.813$ &$-0.700$ &$ 1.683$ \\
      &      &   &$\gamma$ &$6.3f$ &$-0.556$ &$ 1.303$ &$-0.686$ \\
Star &Gra &80 &$\alpha$ &$-93.857$ &$783.889$ &$-0.410$ &$-0.191$ \\
      &      &   &$\beta$  &$ 2.537$ &$-3.040$ &$ 0.035$ &$ 0.846$ \\
      &      &   &$\gamma$ &$6.3f$ &$-0.509$ &$ 1.285$ &$-0.796$ \\
Star &Sil &80 &$\alpha$ &$-330.207$ &$1006.511$ &$-0.428$ &$-0.215$ \\
      &      &   &$\beta$  &$ 3.198$ &$-3.652$ &$-0.547$ &$ 1.632$ \\
      &      &   &$\gamma$ &$6.3f$ &$-0.452$ &$ 0.825$ &$-0.195$ \\

\enddata
\label{tab:interp}
\vspace{-0.25in}
\tablecomments{
   The Spectrum, Dust and $N$ columns indicate the input spectrum (black body
   or stellar atmosphere), the type of dust (Graphitic or silicate) and the
   number of particles in the grain.  The Term column is the coefficient in
   the fitting formula Equation~\ref{eqn:fit} where the four coefficients
   for $\alpha(T_*)$, $\beta(T_*)$ and $\gamma(T_*)$
   are to be summed as $\alpha(T_*)= \sum_{i=0}^3 \alpha_i x^i$ with
   $x =\log_{10}(T_*/1000~\hbox{K})$.  They are valid for $2600 < T_* <30000$~K
   with a typical rms residual compared to the numerical results of $<100$~K
   in the estimate of $T_e$, including the noise in the numerical results.
   Most systematic residuals can be viewed as $\sim 500$~K uncertainties in
   the appropriate $T_*$.
   }
\end{deluxetable}


\begin{thebibliography}{}
\bibitem[\protect\citeauthoryear{Bianchi \& Schneider}{2007}]{Bianchi2007} Bianchi, S., \& Schneider, R.\ 2007, MNRAS, 378, 973               
\bibitem[\protect\citeauthoryear{Castelli \& Kurucz}{2004}]{Castelli2004} Castelli, F., \& Kurucz, R.~L.\ 2004, arXiv:astro-ph/0405087 
\bibitem[\protect\citeauthoryear{Cherchneff \& Dwek}{2009}]{Cherchneff2009} Cherchneff, I., \& Dwek, E.\ 2009, ApJ, 703, 642               
\bibitem[\protect\citeauthoryear{Clayton \& Wickramasinghe}{1976}]{Clayton1976} Clayton, D.~D., \& Wickramasinghe, N.~C.\ 1976, Ap\&SS, 42, 463 
\bibitem[\protect\citeauthoryear{Czerny \& Hryniewicz}{2011}]{Czerny2011} Czerny, B., \& Hryniewicz, K.\ 2011, A\&A, 525, L8 
\bibitem[\protect\citeauthoryear{Davidson}{1987}]{Davidson1987} Davidson, K.\ 1987, ApJ, 317, 760 
\bibitem[\protect\citeauthoryear{Deguchi}{1980}]{Deguchi1980} Deguchi, S.\ 1980, ApJ, 236, 567 
\bibitem[\protect\citeauthoryear{Dessart et al.}{2013}]{Dessart2013} Dessart, L., Hillier, D.~J., Waldman, R., \& Livne, E.\ 2013, MNRAS, 433, 1745 
\bibitem[\protect\citeauthoryear{Desert et al.}{1986}]{Desert1986} Desert, F.~X., Boulanger, F., \& Shore, S.~N.\ 1986, A\&A, 160, 295
\bibitem[\protect\citeauthoryear{Donn \& Nuth}{1985}]{Donn1985} Donn, B., \& Nuth, J.~A.\ 1985, ApJ, 288, 187 
\bibitem[\protect\citeauthoryear{Draine \& Lee}{1984}]{Draine1984} Draine, B.~T., \& Lee, H.~M.\ 1984, ApJ, 285, 89 
\bibitem[\protect\citeauthoryear{Draine \& Anderson}{1985}]{Draine1985} Draine, B.~T., \& Anderson, N.\ 1985, ApJ, 292, 494 
\bibitem[\protect\citeauthoryear{Draine}{1995}]{Draine1995} Draine, B.~T.\ 1995, Ap\&SS, 233, 111 
\bibitem[\protect\citeauthoryear{Dwek}{1988}]{Dwek1988} Dwek, E.\ 1988, ApJ, 329, 814 
\bibitem[\protect\citeauthoryear{Elvis et al.}{2002}]{Elvis2002} Elvis, M., Marengo, M., \& Karovska, M.\ 2002, ApJL, 567, L107 
\bibitem[\protect\citeauthoryear{Ferrarotti \& Gail}{2006}]{Ferrarotti2006} Ferrarotti, A.~S., \& Gail, H.-P.\ 2006, A\&A, 447, 553 
\bibitem[\protect\citeauthoryear{Gail et al.}{1984}]{Gail1984} Gail, H.-P., Keller, R., \& Sedlmayr, E.\ 1984, A\&A, 133, 320               
\bibitem[\protect\citeauthoryear{Gail \& Sedlmayr}{1985}]{Gail1985} Gail, H.-P., \& Sedlmayr, E.\ 1985, A\&A, 148, 183 
\bibitem[\protect\citeauthoryear{Gail \& Sedlmayr}{1988}]{Gail1988} Gail, H.-P., \& Sedlmayr, E.\ 1988, A\&A, 206, 153 
\bibitem[\protect\citeauthoryear{Gail \& Sedlmayr}{2013}]{Gail2013} Gail, H.-P., \& Sedlmayr, E.\ 2013, Physics and Chemistry of Circumstellar Dust Shells, by Hans-Peter Gail , Erwin Sedlmayr, Cambridge, UK: Cambridge University Press, 2013  
\bibitem[\protect\citeauthoryear{Gall et al.}{2011}]{Gall2011} Gall, C., Hjorth, J., \& Andersen, A.~C.\ 2011, A\&ARv, 19, 43 
\bibitem[\protect\citeauthoryear{Gauger et al.}{1990}]{Gauger1990} Gauger, A., Sedlmayr, E., \& Gail, H.-P.\ 1990, A\&A, 235, 345               
\bibitem[\protect\citeauthoryear{Guhathakurta \& Draine}{1989}]{Guhathakurta1989} Guhathakurta, P., \& Draine, B.~T.\ 1989, ApJ, 345, 230 
\bibitem[\protect\citeauthoryear{Gustafsson et al.}{2008}]{Gustafsson2008} Gustafsson, B., Edvardsson, B., Eriksson, K., et al.\ 2008, A\&A, 486, 951 
\bibitem[\protect\citeauthoryear{H{\"o}nig \& Kishimoto}{2010}]{Honig2010} H{\"o}nig, S.~F., \& Kishimoto, M.\ 2010, A\&A, 523, A27 
\bibitem[\protect\citeauthoryear{Humphreys \& Davidson}{1994}]{Humphreys1994} Humphreys, R.~M., \& Davidson, K.\ 1994, PASP, 106, 1025 
\bibitem[\protect\citeauthoryear{Ivezic \& Elitzur}{1997}]{Ivezic1997} Ivezic, Z., \& Elitzur, M.\ 1997, MNRAS, 287, 799
\bibitem[\protect\citeauthoryear{Ivezic et al.}{1999}]{Ivezic1999} Ivezic, Z., Nenkova, M., \& Elitzur, M.\ 1999, User Manual for DUSTY,
  University of Kentucky Internal Report http:\/\/www.pa.uky.edu\/\~moshe\/dusty\/
\bibitem[\protect\citeauthoryear{Jerkstrand et al.}{2012}]{Jerkstrand2012} Jerkstrand, A., Fransson, C., Maguire, K., et al.\ 2012, A\&A, 546, A28 
\bibitem[\protect\citeauthoryear{Johnson et al.}{1993}]{Johnson1993} Johnson, D.~J., Friedlander, M.~W., \& Katz, J.~I.\ 1993, ApJ, 407, 714 
\bibitem[\protect\citeauthoryear{Keith \& Lazzati}{2011}]{Keith2011} Keith, A.~C., \& Lazzati, D.\ 2011, MNRAS, 410, 685               
\bibitem[\protect\citeauthoryear{Knapp}{1985}]{Knapp1985} Knapp, G.~R.\ 1985, ApJ, 293, 273 
\bibitem[Kochanek(2011)]{Kochanek2011} Kochanek, C.~S.\ 2011, ApJ, 743, 73 
\bibitem[\protect\citeauthoryear{Kochanek et al.}{2012}]{Kochanek2012} Kochanek, C.~S., Szczygie{\l}, D.~M., \& Stanek, K.~Z.\ 2012, ApJ, 758, 142 
\bibitem[\protect\citeauthoryear{Kozasa et al.}{1991}]{Kozasa1991} Kozasa, T., Hasegawa, H., \& Nomoto, K.\ 1991, A\&A, 249, 474                
\bibitem[\protect\citeauthoryear{Kwok}{1975}]{Kwok1975} Kwok, S.\ 1975, ApJ, 198, 583 
\bibitem[\protect\citeauthoryear{Lafon \& Berruyer}{1991}]{Lafon1991} Lafon, J.-P.~J., \& Berruyer, N.\ 1991, A\&ARv, 2, 249 
\bibitem[\protect\citeauthoryear{Laor \& Draine}{1993}]{Laor1993} Laor, A., \& Draine, B.~T.\ 1993, ApJ, 402, 441 
\bibitem[\protect\citeauthoryear{Lefevre}{1979}]{Lefevre1979} Lefevre, J.\ 1979, A\&A, 72, 61 
\bibitem[\protect\citeauthoryear{Malloci et al.}{2007}]{Malloci2007} Malloci, G., Joblin, C., \& Mulas, G.\ 2007, Chemical Physics, 332, 353 
\bibitem[\protect\citeauthoryear{Nanni et al.}{2013}]{Nanni2013} Nanni, A., Bressan, A., Marigo, P., \& Girardi, L.\ 2013, MNRAS, 434, 2390 
\bibitem[\protect\citeauthoryear{Netzer \& Elitzur}{1993}]{Netzer1993} Netzer, N., \& Elitzur, M.\ 1993, ApJ, 410, 701 
\bibitem[\protect\citeauthoryear{Nozawa et al.}{2003}]{Nozawa2003} Nozawa, T., Kozasa, T., Umeda, H., Maeda, K., \& Nomoto, K.\ 2003, ApJ, 598, 785 
\bibitem[\protect\citeauthoryear{Ofek et al.}{2014}]{Ofek2014} Ofek, E.~O., Sullivan, M., Shaviv, N.~J., et al.\ 2014, arXiv:1401.5468 
\bibitem[\protect\citeauthoryear{Omont}{1986}]{Omont1986} Omont, A.\ 1986, A\&A, 164, 159 
\bibitem[\protect\citeauthoryear{Rowan-Robinson \& Harris}{1983}]{Rowan1983} Rowan-Robinson, M., \& Harris, S.\ 1983, MNRAS, 202, 797 
\bibitem[\protect\citeauthoryear{Salpeter}{1977}]{Salpeter1977} Salpeter, E.~E.\ 1977, ARA\&A, 15, 267 
\bibitem[\protect\citeauthoryear{Scalo \& Slavsky}{1980}]{Scalo1980} Scalo, J.~M., \& Slavsky, D.~B.\ 1980, ApJL, 239, L73 
\bibitem[\protect\citeauthoryear{Sellgren}{1984}]{Sellgren1984} Sellgren, K.\ 1984, ApJ, 277, 623 
\bibitem[\protect\citeauthoryear{Smith et al.}{2011}]{Smith2011} Smith, N., Li, W., Filippenko, A.~V., \& Chornock, R.\ 2011, MNRAS, 412, 1522
\bibitem[\protect\citeauthoryear{Todini \& Ferrara}{2001}]{Todini2001} Todini, P., \& Ferrara, A.\ 2001, MNRAS, 325, 726               
\bibitem[\protect\citeauthoryear{Ventura et al.}{2014}]{Ventura2014} Ventura, P., Dell'Agli, F., Schneider, R., et al.\ 2014, MNRAS, 439, 977 
\bibitem[\protect\citeauthoryear{Vink}{2012}]{Vink2012} Vink, J.~S.\ 2012, Astrophysics and Space Science Library, 384, 221 
\bibitem[\protect\citeauthoryear{Voit}{1991}]{Voit1991} Voit, G.~M.\ 1991, ApJ, 379, 122                
\bibitem[\protect\citeauthoryear{Voit}{1992}]{Voit1992} Voit, G.~M.\ 1992, MNRAS, 258, 841      
\bibitem[\protect\citeauthoryear{Waxman \& Draine}{2000}]{Waxman2000} Waxman, E., \& Draine, B.~T.\ 2000, ApJ, 537, 796 
\bibitem[\protect\citeauthoryear{W\"olk et al.}{2002}]{Wolk2002} W\"olk, J., Strey, R., Heath, C.H., \& Wyslouzil, B.,  2002, J Chem Phys, 117, 4954
\end{thebibliography}
\end{document}